# Surrogate Structure-Specific Probabilistic Dynamic Responses of Bridge Portfolios using Deep Learning with Partial Information

Chunxiao Ning[1] | Yazhou Xie[1]

**Abstract:**
Predicting region-wide structural responses under seismic shaking is essential for enhancing the effectiveness of earthquake engineering task forces such as earthquake early warning and regional seismic risk and resilience assessments. Existing domain-specific and data-driven approaches, however, lack the capability to provide high-fidelity, structure-specific dynamic response predictions for large-scale structural inventories in a timely manner. To address this gap, this study designed a novel deep learning framework, which integrates heterogeneous ground motion sequences and partial structural information as model inputs, to predict structure-specific, probabilistic dynamic responses of regional structural portfolios. Validation on a portfolio of highway bridges in California demonstrates the model's ability to capture inter-structure response variability by inputting critical and accessible bridge parameters while accounting for uncertainties due to the lack of other information. The results underscore the framework's efficiency and accuracy, paving the way for various advancements in performance-based earthquake engineering and regional-scale seismic decision-making.

## Introduction

Over the past two decades, earthquakes and their secondary disasters have ranked as the deadliest natural hazards globally, accounting for 58% of disaster-related fatalities[1]. Earthquakes caused extensive damage to structures and infrastructure systems in highly urbanized regions, as evidenced by the 2011 earthquake and tsunami in Japan; this hazard resulted in an unprecedented economic loss of $239 billion USD[1]. Earthquake-related research and practice must be addressed at a regional scale[2,3] to account for spatial variations in seismic risk, infrastructure needs, emergency response, structural portfolio impacts, and resource allocation[4,5]. Earthquake-prone, urban regions like California face heightened seismic risks (Fig. 1(a)), threatening to undermine the lifelines of their expansive transportation networks (Fig. 1(b)). Emerging task forces have been established to assess and mitigate the regional impacts of seismic hazards, focusing on earthquake early warning (EEW) systems, regional ground motion (GM) hazard simulations, as well as informed decision-making guided by seismic risk and resilience assessments[6–8].

EEW systems aim to safeguard communities by rapidly detecting earthquake events, estimating the expected shaking intensity, and delivering real-time notifications to the public[9]. The system uses sensors to pick up the initial less damaging P-wave signal and issues alerts before the arrival of the more destructive S-wave (Fig.1(d))[9], which provides a critical window of time to reduce injuries, save lives, and minimize damage to infrastructure[10]. Expected to reduce the number of injuries by more than 50%[11], EEW systems are now operational in many countries such as the USA[12], Japan[13], Mexico[14], and China[15].

Regional GM simulations predict seismic shaking maps against hypothetical earthquake scenarios in the future. It originates from probabilistic seismic hazard analysis (PSHA) that estimates the likelihood and intensity of earthquake-induced ground shaking at different sites by integrating uncertainties and correlations in seismic sources[16], magnitudes, and ground motions[5,17] (Fig.1(e)). Recently, physics-based simulations provided an alternative means to more realistically estimate earthquake-induced ground shaking. They utilize numerical techniques with detailed geological data to explicitly simulate the physical processes of earthquake generation and seismic wave propogation[18]. The simulated synthetic seismograms supplement empirical GM databases and are particularly useful for conditions with limited recorded GMs, such as large-magnitude earthquakes at short source-to-site distances[19,20].

Seismic risk and resilience models evaluate the socio-economic impacts of earthquake hazards on the built environment, enabling stakeholders to better prepare and plan for, absorb, recover from, and more successfully adapt to earthquake events[21] (Fig. 1(f)). The risk/resilience metrics are typically quantified by integrating PSHA results, structural responses, and consequence models to estimate seismic loss and simulate post-earthquake recovery[22].

The effectiveness of these task forces hinges on a critical factor: the ability to reliably predict structural responses under seismic shaking. Current EEW systems operate by estimating seismic intensities at target locations. Their efficacy would be significantly enhanced if structural responses and damage states could be predicted in real-time for alert warnings[9,23,24]. Physics-based simulations generate synthetic GM fields across a region, offering significant advantages for directly predicting dynamic responses and inferring seismic damage[25]. Seismic risk and resilience assessments typically rely on fragility models[26] to estimate structures' probabilities of exceeding various damage states. Fragility models are commonly developed through an analytical approach that performs nonlinear response history analyses (NRHAs) of structures subjected to a selected set of GMs. Reliable dynamic response predictions can enhance EEW systems, leverage results from physics-based GM simulations, and refine fragility models to better support seismic risk and resilience assessments.

Dynamic responses of structures can be obtained through numerical simulation, seismic instrumentation, and shake table testing (Fig. 1(c)). While seismic instrumentation provides real-world measurements of structural responses under seismic shaking[27], its effectiveness is constrained by sparse sensor coverage, high costs, noise contamination, and limited response data under strong earthquakes. Shake table testing has recently been applied to full-scale structural systems[28]; however, most tests remain constrained by table size, the scaling challenge, high cost, and the inability to fully replicate real-world boundary conditions. Numerical simulation provides one viable strategy to predict dynamic responses of structures through high-fidelity finite element models that capture realistic material behaviors, complex component interactions, and nonlinear dynamics under seismic loading[29]. However, numerical simulation is not without challenges in dealing with regional structures. The vast building stock and infrastructure inventory in urban regions makes it impractical to develop high-resolution finite element models for every structure, given the sophisticated variations in geometry, material properties, and design details. Most existing structural and infrastructure databases fail to fully capture such variations, typically providing only general information, such as construction years and broad geometric parameters (e.g., the number of spans for bridges or stories for buildings[30–32]). This lack of detail hinders the extraction of all necessary data required to develop high-fidelity simulation models for the entire structural portfolio. Even if such digitized assets with complete structural information are readily available, performing NRHAs for all structures under all possible earthquake scenarios is computationally infeasible. In regional seismic risk/resilient assessment, a crucial step to ensure fidelity is to refine fragility models by transitioning from archetype-based class fragilities[33] to structure-specific parameterized fragilities[34,35]. However, developing such refined fragility models requires an extensive number of NRHAs, as exemplified by the 320,000 analyses conducted by Goda and Atkinson[34] for wood-frame houses in southwestern British Columbia and the 44,010 analyses performed by Abarca et al.[36] for 163 bridges in Italy.

[1]Department of Civil Engineering, McGill University, Montreal, QC, Canada. Email: tim.xie@mcgill.ca

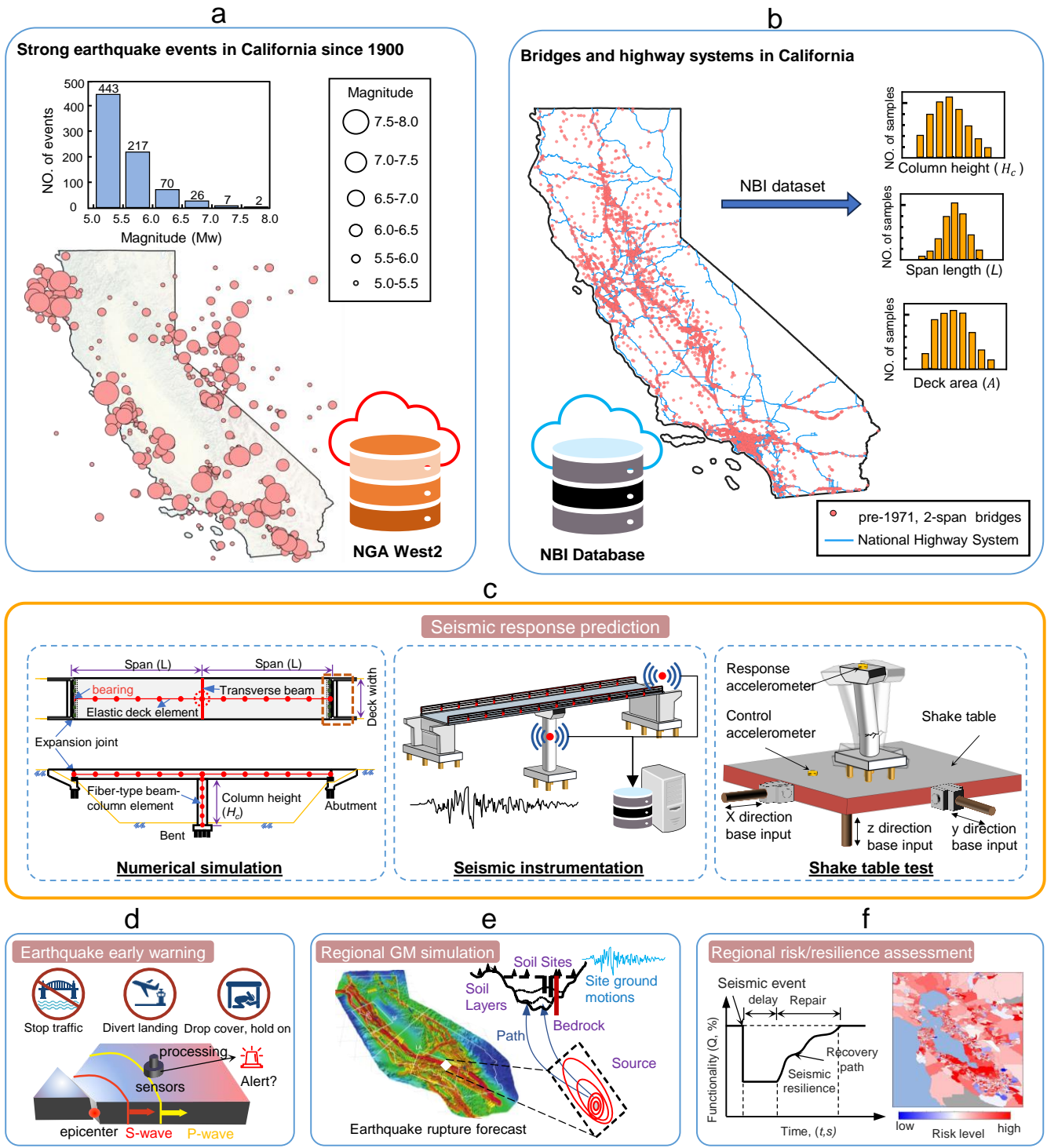

**FIGURE 1. Seismic response prediction supports various task forces in earthquake engineering.** (**a**) Earthquake events in California with magnitude greater than 5.0 from 1900 to 2024. GM records at various locations are available in the NGA West2 database[46]. (**b**) Highway systems and pre-1971, two-span bridges in California. Certain geometric parameters of these bridges are available in the NBI database[32]. (**c**) Three approaches for dynamic response predictions of bridges under seismic loading. (**d**) Earthquake early warning system. (**e**) Regional ground motion simulation through probabilistic seismic hazard analysis. Earthquake rupture forecast adopts the UCERF3 model[16] (**f**) Regional seismic risk and resilience assessments for the functionality recovery curve and seismic risk map[5].

Deep learning techniques have recently emerged as effective surrogates to replace high-fidelity finite element models for efficient predictions of complete structural response histories during seismic shaking. Various types of neural networks, such as convolutional neural networks[37] and recurrent neural networks[38], have been developed to predict seismic response histories of buildings and bridges. These models are typically trained on a specific structure using GM sequences as the sole input, yet they fail to generalize the predictions across different structures across a region. A few studies[39,40] have extended these frameworks to predict seismic responses of different homogeneous structures by integrating structural parameters into the model inputs. However, these studies fail to address the inventory data challenges mentioned above. On one hand, they often require sophisticated structural parameters (e.g., natural periods) as model inputs, which cannot be conveniently extracted from existing inventory databases. On the other hand, they treat the prediction process as deterministic, overlooking the fact that the structural parameters used in deep learning are frequently incomplete for high-fidelity response predictions. Surrogate response predictions must also take into account uncertainties due to the incompleteness of structural parameters as model inputs.

This study develops a novel Structure Portfolio Response Prediction Network (SPR-Net) to predict the distributions of dynamic response histories using GM and partially available structural parameters as model inputs. This approach captures inter-structure variations using available and critical structural parameters while treating missing parameters as sources of uncertainty. The SPR-Net integrates the fully connected layers, dilated casual convolutional layers, recurrent layers, and gate mechanisms to effectively process heterogeneous data inputs, where a hybrid loss function is devised to enable the probabilistic response prediction. A feature selection module is also integrated to identify critical structural parameters as incomplete input information through a transfer learning[41] mechanism. The efficacy and accuracy of this method are demonstrated against a portfolio of highway bridges in the state of California, as shown in Fig. 1(b).

## Results

### Structural Portfolio Response Prediction Network (SPR-Net)

The model architecture of the developed SPR-Net is shown in Fig. 2(a). It integrates fully connected layers, dilated casual convolutional layers, recurrent layers, and gate mechanisms to effectively process heterogeneous data inputs and make probabilistic seismic response predictions of structural portfolios. The SPR-Net is designed with two distinct channels: the dilated causal convolutional layer channel[42,43] for processing GM sequences as time-series data, and a fully connected layer channel for handling structural parameters as vector inputs. The heterogeneous inputs are then fused through the gate mechanism, where essential features are extracted and combined through the skip connection and residual connection mechanisms. Depending on whether the complete set of structural parameters is available or not, the SPR-Net can be adapted to make deterministic or probabilistic response predictions. Fig. 2(c)-(e) indicates the workflow to establish the probabilistic SPR-Net where partial structural information is available. The workflow involves a pre-training of the deterministic SPR-Net, the selection of key structural features, and transfer learning of the probabilistic SPR-Net. The deterministic SPR-Net is first trained using results from high-fidelity numerical simulations that incorporate all structural parameters. The feature selection module then computes the Shapley values[44] of all structural parameters and identifies the important ones that (1) contribute most to the response predictions and (2) can be easily extracted from existing inventory databases. With the selected key structural parameters and GM data, the transfer learning technique is used to train the probabilistic SPR-Net model in Fig. 2(e), which freezes the GM processing channel and trains other channels. The probabilistic SPR-Net model predicts the mean and standard deviation of the response history at each time step following a lognormal distribution, as shown in Fig. 2(b). The loss function for the deterministic SPR-Net is the mean squared error between ground truth and response prediction. The loss function for the probabilistic SPR-Net is composed of the data loss, which is the mean squared error between the ground truth and the predicted median, and the uncertainty loss that regulates the negative log-likelihood of producing the observations from the predicted median and standard deviation. More details regarding the SPR-Net are provided in the method section.

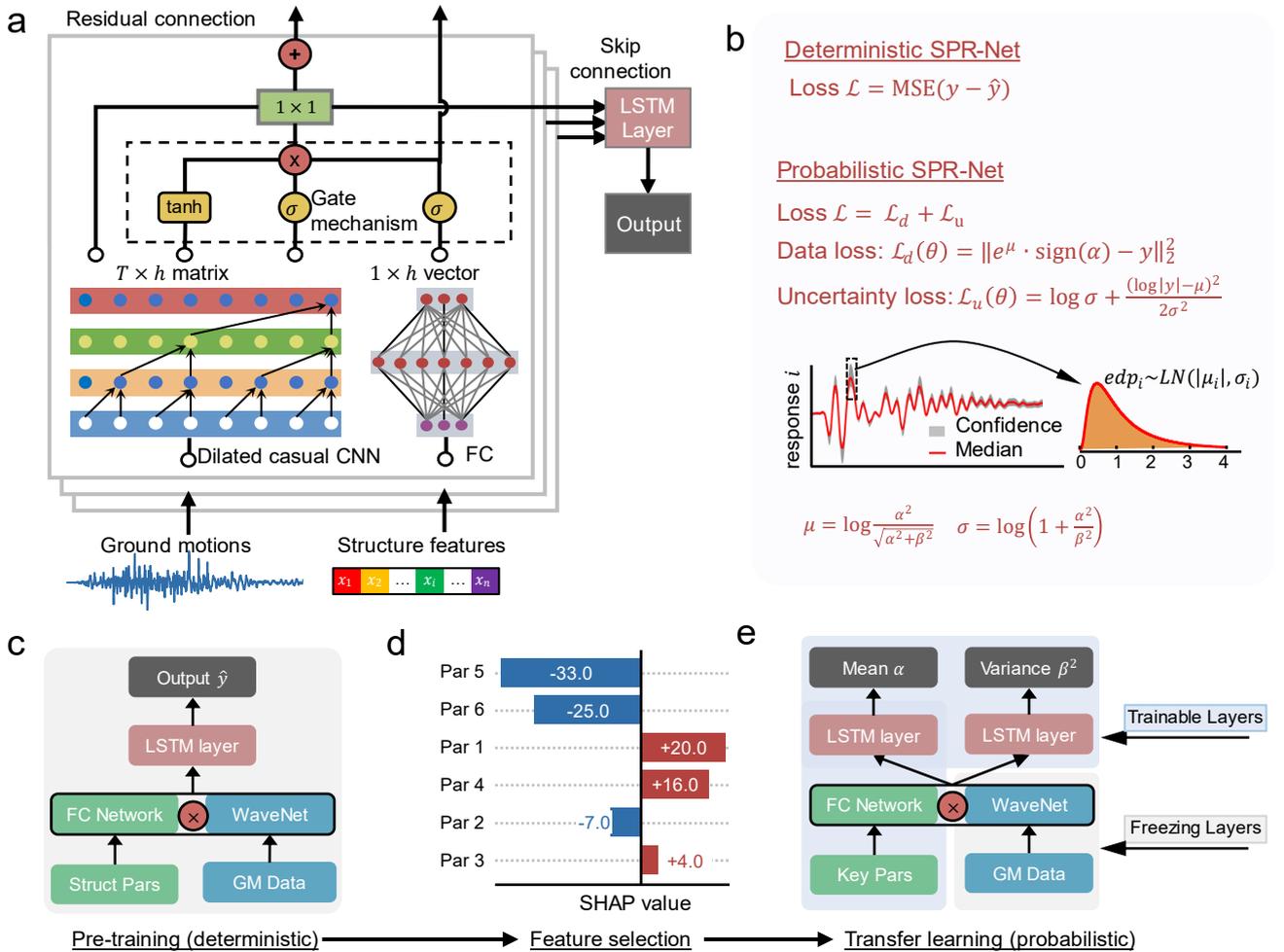

**FIGURE 2. Schematic framework of the SPR-Net for the seismic response prediction of structure portfolios.** (**a**) The architecture of the designed SPR-Net is composed of fully connected layers processing structure features, and dilated casual convolutional layers processing GM sequence. The features from structure and GM are merged through the gate mechanism. (**b**) Loss functions for training the SPR-Net. The loss function for the probabilistic prediction is composed of uncertainty loss $\mathcal{L}_u$ and data loss $\mathcal{L}_d$. (**c**) The deterministic SPR-Net with all structure features as inputs and dynamic response history as the output. (**d**) Feature selection based on Shapley values and data accessibility. (**e**) The probabilistic SPR-Net with selected key structural parameters and GM sequence as input, and the mean and variance of the dynamic response history as outputs. The model is transferred from the deterministic SPR-Net by freezing the module processing the GM data and fine turning other modules. The distribution of seismic response at each time step is considered to follow a lognormal distribution, and the probabilistic SPR-Net predicts the mean $\alpha$ and variance $\beta$ of the distribution.

## Development of Benchmark Case-study Dataset

The case study predicts seismic responses of a highway bridge portfolio in California, USA. The selected bridge class is the two-span, single-column, continuous concrete box-girder bridges with seat-type abutments constructed before 1971. The pre-1971 two-span bridges take a significant portion of the bridge network in California, as illustrated in Fig. 1(b)[33,45]. The National Bridge Inventory (NBI)[32] provides basic bridge parameters, which are combined with common design practice values to form the partial information for developing the probabilistic SPR-Net. Conversely, detailed bridge parameters are extracted from a comprehensive review of design drawings for this bridge class in California[33]. A total of 650 GM records are selected from the NGA-West2 database[46], comprising 221 records from earthquake events in California and 429 from events outside California. All ground motions are recorded from shallow crustal earthquakes in active tectonic regions, which resemble the seismicity in California. Each GM was scaled twice with a random scaling factor ranging from 1 to 3, resulting in a total suite of 1950 GM records. Using the statistical distribution of each detailed bridge parameter, 1,950 bridge samples are generated through the Latin hypercube sampling method. These bridge samples are then randomly paired with the GMs to develop high-fidelity finite element models and perform NRHA, resulting in the case-study dataset comprising GM inputs, detailed bridge parameters, and dynamic response outputs for different bridge components, such as column and bearing. This dataset is divided into subsets with 900 instances for training, 240 for validation, and 810 for testing, ensuring no overlap in between. Further details on the detailed bridge parameters, high-fidelity numerical model, and GM dataset are provided in the Supplementary Material.

## Pre-Training and Feature Selection

The deterministic SPR-Net is trained against response histories of the column drift ratio $\Delta_C$, column lateral force $F_C$, bearing displacement $\gamma$, and bearing lateral force $F_b$, using time-series inputs from GM records and vector inputs from 15 detailed bridge parameters, as shown in Figs. S1–S2 in the Supplementary Material. Explanations of these 15 bridge parameters are provided in the Method Section. The SHapley Additive exPlanations (SHAP)[44] technique is then employed to analyze the training process and rank the significance of each bridge parameter. Fig. 3 indicates the top 10 most influential bridge features concerning the maximum column drift and bearing deformation. The results demonstrate the model's ability to capture the underlying physics of how bridge parameters affect seismic responses. For example, column drift and bearing displacement would be increased if the bridge has a lower stiffness due to the slender column ($\lambda$), larger pounding gap ($\delta$), and longer span ($L$). Additional insights into the influence of bridge parameters on other response metrics are provided in Fig. S8.

While Shapley values offer a basis for ranking and selecting influential bridge features, one should also consider whether their values can be extracted conveniently, either from existing inventory databases (e.g., the NBI) or inferred through typical design practices. The study takes bridge geometric parameters, namely column slenderness ($\lambda$), span length ($L$), and abutment seat gap ($\delta$), as available partial information to develop the probabilistic SPR-Net. Conversely, other bridge parameters, such as column reinforcement ratio ($\rho_s$), system damping ratio ($\xi$), bearing stiffness ($k_b$), are excluded due to their negligible impact or the difficulty of extracting their values without testing measurements or reviewing detailed design drawings.

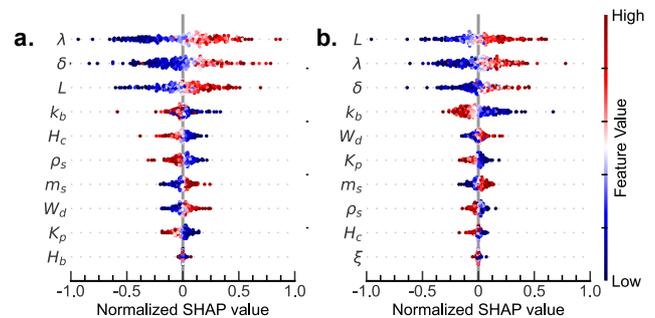

**FIGURE 3.** Feature importance by Shapley value against (**a**) the maximum column drift and (**b**) the maximum bearing displacement.

## Seismic Response Prediction through Probabilistic SPR-Net

The structural parameter processing channel and the output layers of the deterministic SPR-Net are modified from Fig. 2(c) to Fig. 2(e). This transfer learning process re-trains the model using only key structural parameters towards probabilistic response predictions. In particular, the GM processing channel is frozen while other channels are trainable. For comparison purposes, the long short-term memory neural network (LSTM)[38,43] is also trained as a base model with GM sequence as the sole input. Fig. 4 presents comparisons of the cumulative distribution functions (CDFs) from the testing dataset for column and bearing responses, including peak drift/deformation, peak lateral force, and dissipated energy. Results from the probabilistic SPR-Net are provided as the median prediction and those with one standard deviation confidence intervals. Fig. 4 shows the superior performance of the proposed SPR-Net considering both accuracy and uncertainty quantification: its predicted median response aligns well with the ground truth CDF against most response metrics. Conversely, the LSTM shows notable underestimations in capturing large responses. Taking 84[th] CDF percentiles of column responses as the example, the LSTM yields relative errors between 11%-25%, whereas the median predictions from the SPR-Net are much closer, with relative errors at 1%-7%. The probabilistic SPR-Net also captures its uncertainties in response predictions; the one standard deviation results show a reasonable interval that encompasses the ground truth data.

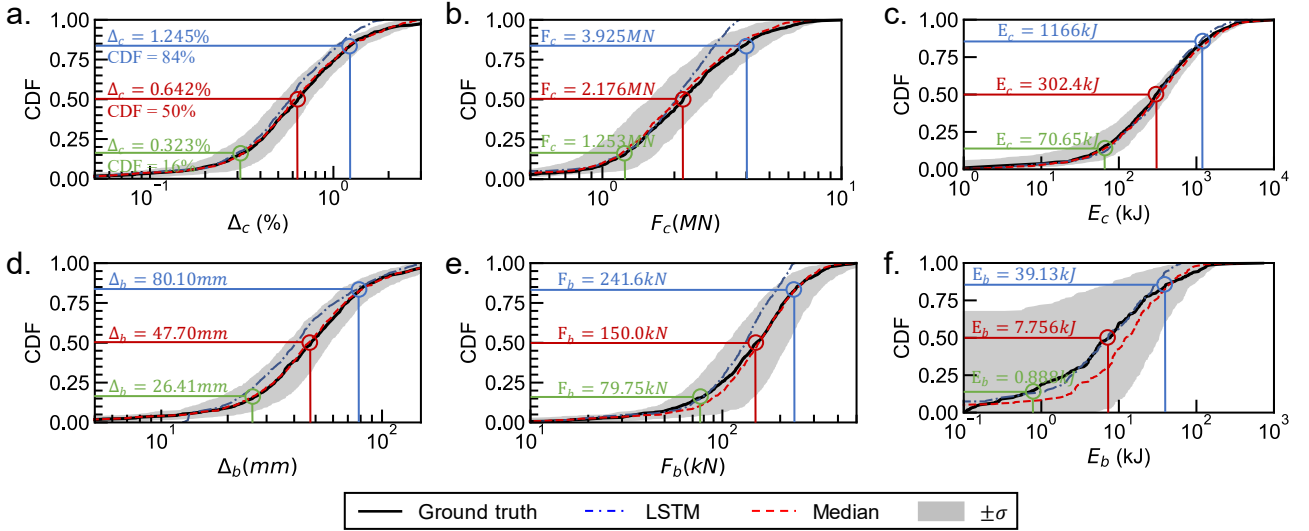

**FIGURE 4.** CDF comparisons of different response metrics from LSTM and SPR-Net. (a) peak column drift ratio, (b) peak column lateral force, (c) dissipated energy of column, (d) peak bearing displacement, (e) peak bearing lateral force, and (f) dissipated energy of bearing. 16[th], 50[th] and 84[th] percentile CDF values are listed in each figure.

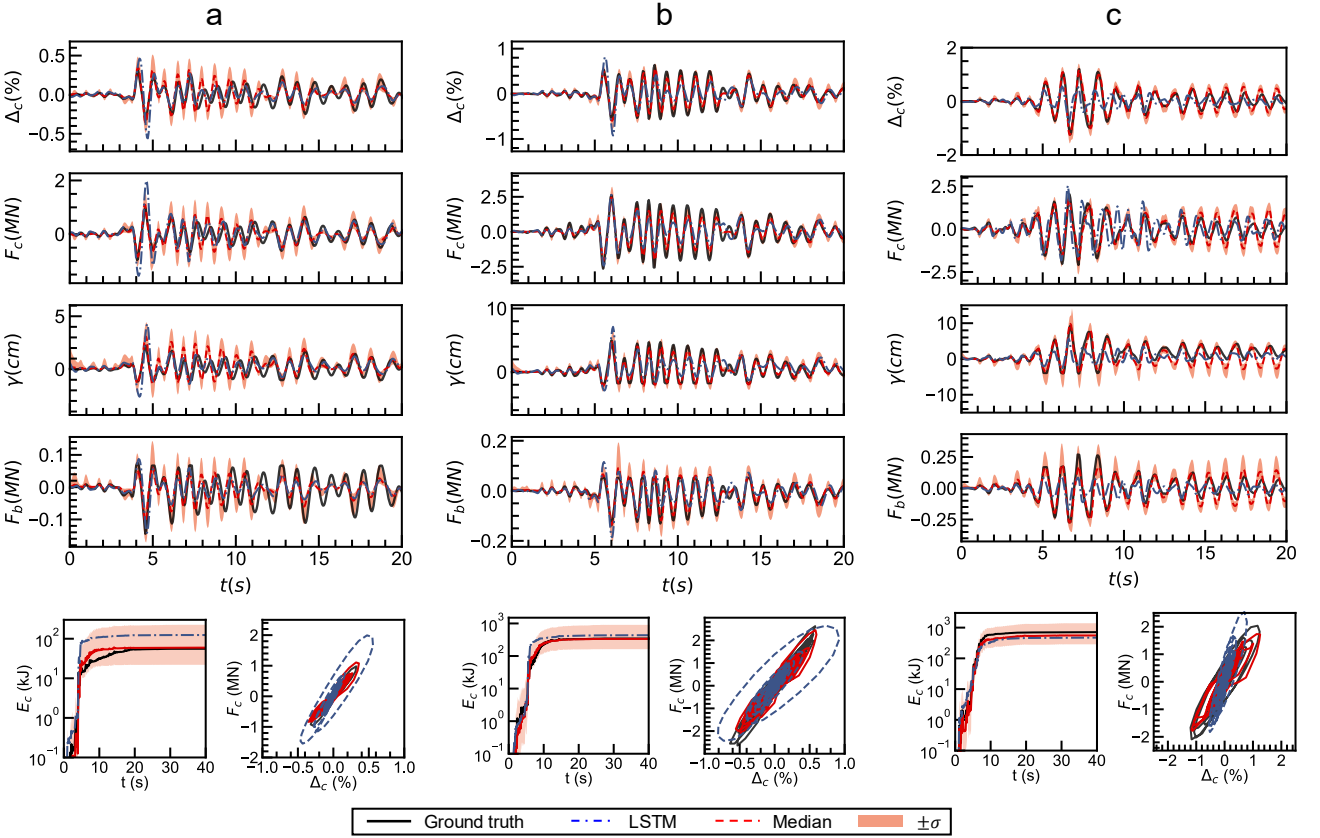

**FIGURE 5.** Time series response comparisons for column drift ratios at (a) 16[th], (b) 50[th], and (c) 84[th] CDF percentiles.

Fig. 5 further presents time-series response comparisons for three cases selected from the test dataset. The selected cases correspond to the results that have peak column drift ratios at the 16[th], 50[th], and 84[th] CDF percentiles. The median response predictions from the SPR-Net closely align with the ground truth data concerning both column drift ($\Delta_c$) and bearing displacement ($\gamma$), demonstrating its ability to capture the amplitude, frequency content, and phase information of the entire dynamic response histories of bridge components. While slight discrepancies are observed in the predicted histories of lateral force demands on columns ($F_c$) and bearings ($F_b$), the SPR-Net model predicts the shaded confidence intervals that encompass the ground truth responses in all cases. The confidence intervals for column drift ratios and bearing displacements are narrow, indicating the strong capability of the SPR-Net model in predicting bridge displacements with a small uncertainty. In contrast, the LSTM model shows larger deviations from the ground truth data, particularly in capturing peak responses and overall history shapes. For the 16$^{th}$ percentile case (Fig. 5(a)), the peak column drift ratio is 0.32% from the numerical simulation; the SPR-Net predicts a median of 0.36% with interval bounds at 0.27%-0.51%. However, the LSTM model predicts an unrealistic peak value of 0.56%. For the 50$^{th}$ and 84$^{th}$ percentile cases, the SPR-Net demonstrates near-perfect predictions, with the predicted medians and narrow confidence intervals closely overlapping the ground truth response curves. Specifically, the ground truth for the 84$^{th}$ percentile case has a peak column drift of 1.24%, which is consistent with the SPR-Net predictions at the median of 1.25% and confidence bounds of 1.00%-1.57%. In contrast, the LSTM model underestimates the value, predicting a peak column drift at 0.79%, an error of approximately 36%.

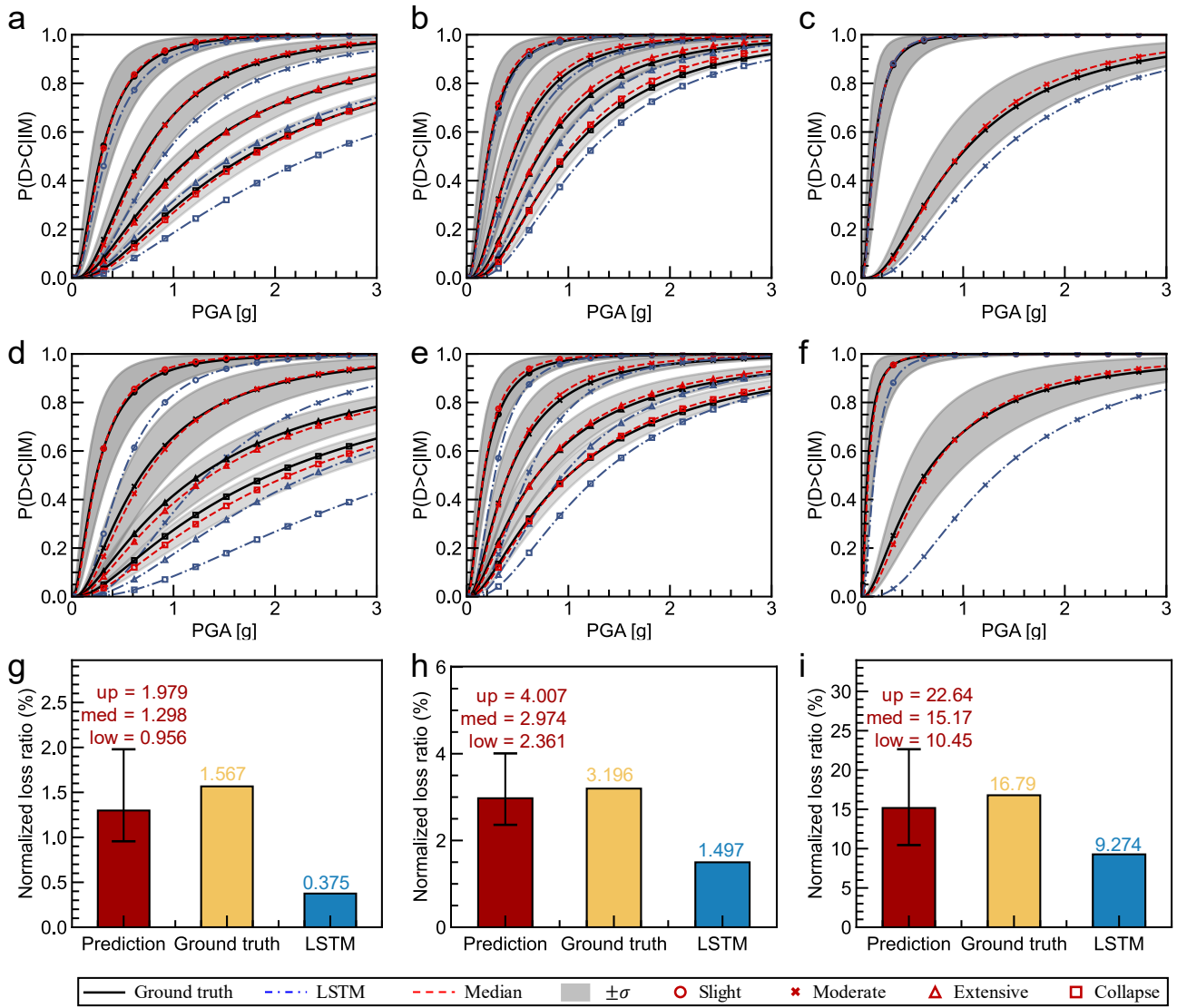

**FIGURE 6.** Comparisons of seismic fragility curves and seismic loss estimations. **Class fragility:** (a) column using drift ductility (b) column using the Park-Ang index (c) bearing using peak displacement; **Fragility for the soft bridge:** (d) column using drift ductility (e) column using the Park-Ang index (f) bearing using peak displacement (g) column loss ratio using drift ductility (h) column loss ratio using the Park-Ang index and (i) bearing loss ratio using peak displacement.

Fig. 5 also compares predictions on the cumulative dissipated energy ($E_C$) and force-displacement hysteretic loops of bridge columns. The LSTM model overestimates the column drifts and energies for the 16$^{th}$ and 50$^{th}$ CDF percentile cases while underestimating the values for the 84$^{th}$ percentile case. Conversely, the SPR-Net reliably predicts the column's dissipated seismic energy, with matching hysteretic curves for all three cases against the ground truth data. Overall, the proposed probabilistic SPR-Net model significantly outperforms LSTM in terms of both accuracy and uncertainty quantification. Comparisons on dynamic response predictions demonstrate that SPR-Net is a more reliable and robust model than LSTM.

## Class and Bridge-Specific Seismic Fragility Assessments and Loss Estimations

As mentioned, one application of dynamic response prediction of bridge components is to support the development of their fragility models and conduct seismic risk/resilience assessments. The *Cloud analysis* approach[47] detailed in the Methods Section is utilized to develop bridge fragility models for (1) the entire archetype class (as a low-resolution fragility representation of the entire class where intra-class bridge variance is treated as uncertainty) and (2) individual bridges, where bridge parameters are considered model inputs for differentiating bridge performance towards high-resolution bridge-specific fragility. Fragility models can be developed using different engineering demand parameters (EDPs), including peak values, such as the drift ductility for the column and lateral displacement for the bearing, as well as the energy-dependent index that relates seismic damage to the entire response histories of bridge components. On such EDP is the Park-Ang damage index[48] for the bridge column.

Capacity models for these EDPs are listed in Table S2 considering four damage states for columns and two for bearings.

Fig. 6(a)-(c) compares class fragility curves of bridge column and bearing derived from ground truth data, LSTM predictions, and SPR-Net predictions. The fragilities derived from the median SPR-Net predictions align well with the ground truth curves across all damage states and EDPs. The probabilistic predictions from SPR-Net also provide uncertainty measurements for fragility bounds that consistently encompass the ground truth fragility. In contrast, LSTM significantly underestimates the class fragility, especially for higher damage states. For example, for column collapse fragility using drift ductility, the peak ground acceleration (PGA) at a 50% probability of exceedance is 1.70g from the ground truth data. The SPR-Net predicts a median PGA of 1.73g within a range between 1.55g-1.90g. However, the LSTM prediction is 2.40g, representing a 41.2% relative error.

Bridge-specific fragility models are developed for three representative structures, with their natural periods ranked at the 10[th], 50[th], and 90[th] CDF percentiles in the test dataset, representing stiff, medium, and soft bridges, respectively. Fig. 6(d)-(i) illustrates the fragility results for the soft bridge that has a span length $L = 36.58\,m$, column slenderness ratio $\lambda = 6.25$ and abutment seat gap length $\delta = 30.48\,mm$, resulting in a bridge natural period of 1.03 seconds. With other bridge parameters treated as unknowns, probabilistic seismic responses are predicted using the SPR-Net. Fragilities for stiff and medium bridges are also developed similarly and are provided in the Supplementary Material. Fig. 6(d)-(f) depict the comparisons of seismic fragility curves for column and bearing from the soft bridge. Similar observations on the superior performance of the SPR-Net can be found as those shown in Fig. 6(a)-(c), including a matching median fragility and consistent bounds from the probabilistic predictions. The LSTM model, however, significantly underestimates the column and bearing fragility for the soft bridge at higher damage states.

Fig. 6(g)-(i) further compare the seismic repair cost ratios of the column and bearing for the soft bridge. The bridge is considered to be located in Los Angeles, California (33.98, -117.34), where seismic hazard data is available through the USGS unified hazard tool[49], as shown in Fig. S9. No matter which EDP is considered for the column, Fig. 6(g) and 6(h) demonstrate that the SPR-Net predicts consistent normalized column loss ratios against the ground truth value. For instance, the drift-ductility-based results indicate a column loss ratio of 1.57% from the ground truth, which is well captured through the SPR-Net predictions with a median of 1.30% and a probabilistic range between 0.96-1.98%. In contrast, the LSTM model offers a single-point prediction that is substantially lower in estimating column repair costs (i.e., 0.38% from drift-ductility-based estimates). Similar observations can be found for the seismic loss ratios of bearings, as shown in Fig. 6(i).

## Discussion

We develop a novel SPR-Net deep learning model to probabilistically predict the dynamic responses of regionwide bridge portfolios that vary in geometry, material properties, and design details. The SPR-Net model integrates fully connected layers to process bridge features, dilated casual convolutional layers to process GM data, gate mechanisms to infuse these two heterogeneous data inputs, and recurrent layers to predict the temporal patterns embedded in seismic responses of bridge components. The model deals with partial bridge information through transfer learning that fine-tunes the structural parameter processing channel. In particular, span length ($L$), column slenderness ($\lambda$), and abutment seat gap length ($\delta$) are selected as key bridge features because of their high Shapley ranking and the real-world availability of their values. Other bridge parameters, either not influential or not easily accessible, were treated as sources of uncertainty for probabilistic response predictions.

In contrast to the LSTM model which shows substantial discrepancies in predicting dynamic responses of the case-study bridge class, the SPR-Net incorporates partial bridge features with more reliable predictions against different metrics, ranging from response CDF distributions, individual response histories, archetype-base class fragilities, bridge specific fragilities, and seismic loss estimations (more results are provided in the Supplementary Material). In particular, integrating structural features into the SPR-Net enables a physics-consistent characterization of fragility and loss for individual bridges (Fig. S10-S11). The LSTM model overshadows this level of resolution, resulting in underestimated losses for the soft bridge and overestimated values for the stiff one (Fig. S11).

The proposed SPR-Net model has significant potential as an efficient surrogate for addressing different challenging tasks in earthquake engineering. By inputting real-time ground motion data from EEW sensors, the model can instantaneously predict the dynamic responses of bridge portfolios, thereby overcoming a critical limitation of EEW: the lack of engineering-specific metrics to support decision-making related to the triggering of alerts for various end users[9]. The SPR-Net model achieves a balance between high-resolution response predictions against critical bridge attributes and uncertainty quantifications for the lack of information of other parameters. This adaptability allows the model to function as a viable down-streaming tool to deal with seismic shaking maps predicted through regional GM simulations. Moreover, the probabilistic response predictions generated by SPR-Net can be seamlessly integrated into analytical frameworks for regional seismic risk and resilience assessments. These integrations can occur either as a direct module for predicting seismic response and damage or as an intermediate module for developing reliable archetype-based fragility models or bridge-specific fragility models (both presented in the current study). The latter can be tailored to account for variations in fragility parameters across individual bridges within the inventory. Such advanced response and fragility models are poised to refine regional risk and resilience metrics, enhancing their applicability for post-earthquake emergency response, long-term transportation network planning, and seismic retrofitting of high-risk bridge structures. These advancements hold significant promise for improving decision-making processes and resource allocation in the aftermath of seismic events.

The developed SPR-Net is not without limitations. Its training and testing data come from nonlinear dynamic response analyses against numerical simulations of high-fidelity finite element models. The applicability of the SPR-Net remains constrained to the simulation settings (e.g., the bridge class, modeling approach, numerical solver, and GM characteristics), as getting a similar amount of data from experimental testing and field measurement still remains infeasible in the present. Previous studies have validated the numerical models against experimental and field tests of different bridge components, including columns[48], bearings[50], shear keys[51], abutment backfills[52], pounding gaps[53], etc. To further enhance the robustness of SPR-Net, a viable strategy is to explicitly take into account the modeling uncertainty, namely the extent to which numerical models represent real-world structures, against the entire case-study bridge class. Such uncertainty quantification requires the availability of many more strong GM records and reliable strong-response measurements of bridges obtained from seismic instrumentation during large earthquake events.

## Methods

**Numerical simulation to develop the case-study dataset**

The finite element model discretizes a structure into a large number of elements for assembling large-dimension system matrices to solve the following equations of motion under seismic loading:

$$M\ddot{x} + C\dot{x} + Kx = -MIa_g \qquad (1)$$

where $x = x(t) \in \mathbb{R}^{1 \times N}$ is the displacement responses of the system with a dimension of $N$, and $t \in [0, T]$ denotes time. $\dot{x}$ and $\ddot{x}$ are structural velocity and acceleration responses. $a_g = a_g(t) \in \mathbb{R}^{1 \times N}$ represents the acceleration time history of a GM. $M, C, K \in \mathbb{R}^{N \times N}$ are the mass, damping, and stiffness matrices of the structure. Values in $C$ and $K$ are functions of structural responses when entering into the nonlinear range. Structural responses at each time step are obtained by solving Eq. (1) through a time-stepping iterative integration process, requiring the computations of inverse matrices such as $K$ at each iteration which takes a lot of computational time.

We develop finite element models for a portfolio of two-span, single-column, continuous concrete box-girder bridges designed before 1971 in California, United States[33,45]. The bridges vary in geometry, material properties, supporting soil conditions, and design details. Table S1 summarizes the statistical distribution of each bridge parameter based on an extensive plan review of the bridge class to capture different sources of uncertainties[33]. We sample 1950 bridges using these data parameters and couple each bridge sample with each set of bi-directional GMs selected and scaled from the NGA West-2 database (Fig. S1). The GM sequence is also trimmed to have the same length of 60 seconds with a time step of 0.05 second[37]. For each bridge-GM sample, we utilize the high-fidelity modeling scheme illustrated in Fig. S3 to build the finite element model for conducting the dynamic response analyses. Each bridge model can be characterized through 15 parameters, including span length $L$, deck width $W_D$, column height $H_C$, slenderness ratio $\lambda$ (height over diameter), reinforcement ratio $\rho_s$ and concrete strength $f_c$, backfill height $H_b$, abutment pile capacity $K_{pa}$, column foundation translational stiffness $K_{ft}$ and rotational stiffness $K_{fr}$, bearing stiffness $k_b$ and friction coefficient $\mu_b$, abutment longitudinal seat gap $\delta$, mass multiplier $m_s$, and damping coefficient $\xi$. Statistical distributions of these parameters for the sampled bridges are shown in Fig. S2.

## SPR-Net deep learning framework

The proposed framework for developing the SPR-Net is shown in Fig. 2. It includes model pretraining, feature selection, and transfer learning for achieving probabilistic predictions. The SPR-Net consists of the fully connected channel, dilated casual convolutional channel[42], recurrent layers, gated mechanisms, skip connection, and residual connections. Structural parameters and GM data are processed separately through fully connected and convolutional channels. The casual channel constrains the temporal dependency on the time span where the current response is only influenced by previous GM records and independent of future information. The dilated convolution ensures the size of the receptive field expands exponentially without a significant increase in computation burden. The gate mechanism[42] is employed to (1) filter important information in the predicted sequence and (2) infuse features extracted from both fully connected and convolutional channels. The gate mechanism has the following mathematical representation.

$$z_{k+1} = tanh(W_{f,k} * z_k) \odot \sigma(W_{g,k} * z_k) \odot \sigma(W_{k,c} b_k) \qquad (2)$$

where $*$ and $\odot$ denote convolution operation and element-wise multiplication. $\sigma(\cdot)$ is a sigmoid activation function, $k$ denotes the layer index, $f, g$ and $c$ denotes filter, gate, and fully connected layers, separately. $b_k$ and $z_k$ are vector and sequence features from inputs. $W$ is a trainable weight for filter, gate, or fully connected layers. The residual connection overcomes the vanishing/exploding gradient and degradation problems associated with deep neural networks. It enables a deep architecture and shortens the error propagation path for accelerated training. The skip connection plays a similar role as the residual connection but directly connects the output of each hidden layer to the output layer; it uses recurrent cells to further improve the modeling of temporal dependency for long-time series data.

The model pretraining is first carried out by feeding into the dataset that includes the GM data and all structural parameters, which enables the SPR-Net to achieve deterministic response predictions. The Shapley values[44] are then computed to rank structural parameters according to their feature importance in affecting the model prediction. Key features with easily accessible values in practice are selected as partial structural information for developing the probabilistic SPR-Net, while less critical or difficult-to-extract structural parameters are excluded, contributing to the prediction uncertainties. The SPR-Net considers the absolute values of seismic responses at each time stamp following a lognormal distribution. To achieve probabilistic predictions, the output layers are modified to have two separate recurrent layers predicting both the mean ($\alpha$) and variance ($\beta^2$) of dynamic responses at each time step. Thus, the corresponding Gaussian distribution can be parameterized as

$$\mu = log\,\alpha^2 - log\,\sqrt{\alpha^2 + \beta^2} \qquad (3)$$

$$\sigma = log\left(1 + \frac{\alpha^2}{\beta^2}\right) \qquad (4)$$

Therefore, the predicted response $x_i$ at time stamp $t = i$ can be represented by

$$x_i = exp(\mu_i + \sigma_i \cdot \varphi) \cdot sign(\alpha_i) \qquad (5)$$

where $\varphi$ is the standard normal random variable and $sign(\cdot)$ represents the sign function.

## SPR-Net network architecture

The SPR-Net consists of 12 neural network layers. Each fully connected layer contains 4 hidden units; the convolutional layers have 16 filters and a dilated width with a multiplier of 2. LSTM layers are employed to process the skip connection from previous layers before the final prediction. Fifteen bridge parameters are used as a complete set of inputs for the deterministic response prediction. For the probabilistic SPR-Net, two branches of LSTM layers are used to predict the mean and standard deviation simultaneously. Additionally, the fully connected layer channel is adjusted to have an input size of three for processing the selected three bridge parameters.

The loss function for the deterministic SPR-Net is the mean squared error (MSE), as presented in Eq. (7). In contrast, the probabilistic SPR-Net adopts a loss function that combines two loss terms, namely the negative log-likelihood which quantifies the prediction uncertainty and the MSE loss for the median response prediction, as shown in Eqs. (6) and (7), respectively.

$$\mathcal{L}_u(\theta) = log\,\sigma + \frac{(log|y| - \mu)^2}{2\sigma^2} \qquad (6)$$

$$\mathcal{L}_d(\theta) = \|exp(\mu) \cdot sign(\alpha) - y\|_2^2 \qquad (7)$$

The SPR-Net model is trained on the training dataset using a batch size of 180 for 1000 epochs under the learning rate of $1 \times 10^{-4}$ for the Adam optimizer[54]. The loss functions for both training and validation datasets become stable after 500 epochs (Fig. S4).

## Shapley additive explanations

The SHAP algorithm[44] explains the prediction process of a machine learning model by identifying and ranking the importance of each input feature. It ranks input features by comparing model predictions with and without each feature, treating all features as potential contributors. For each prediction, the Shapley value of each feature can be computed as below:

$$y_i = y_{base} + \sum_{j=1}^{n} f(x_{ij}) \qquad (8)$$

where $y_i$ is the predicted value of the $i$th sample, $y_{base}$ is the base value of the prediction across the entire database, $n$ is the total number of features, and $x_{ij}$ denotes the $j$th feature. The Shapley value of the $j$th feature, $f(x_{ij})$, can be expressed as:

$$f(x_{ij}) = \sum_{s \in S_j} \omega(|s|)[v(s \cup \{j\}) - v(s)] \qquad (9)$$

where $S_j$ is the set of all features excluding the $j$th feature, $s$ represents any subset of $S_j$, $|s|$ is the size of that subset, $v(s)$ is the prediction value considering only the features in subset $s$, $v(s \cup \{j\})$ is the prediction value when the $j$th feature is also included, the weight $\omega(|s/)$ is defined by:

$$\omega(s) = s!\,(n - s - 1)!\,n! \qquad (10)$$

The Shapley value systematically quantifies the individual contribution of each feature to the entire model predictions, which enables the ranking and selection of important features.

## Transfer learning for probabilistic predictions

Transfer learning is applied to train the probabilistic SPR-Net using previously trained hyperparameters as a warm start. It freezes the dilated convolutional layers as the base model for processing the GM data, whereas the weights for fully connected layers are re-trained against the selected three bridge parameters. The models are trained by 1000 epochs with a learning rate of $1 \times 10^{-4}$ for the Adam optimizer; the loss functions on both the training and validation datasets become stable after 500 epochs.

## Long short-term memory model

For comparison purposes, the reference LSTM model is also designed to have 12 neural network layers and 32 filters, followed by 2 layers of 1D convolutional layers. The loss function is adopted as the mean squared error, the same as the deterministic SPR-Net model. The LSTM model is trained using a batch size of 180 for 1000 epochs under the learning rate of $1 \times 10^{-3}$ for the Adam optimizer, and the model with the least loss on the validation set is used for comparisons. The training, validation, and testing data sets for the LSTM model are the same as the SPR-Net model.

## *Cloud* analysis and seismic fragility modeling

Seismic fragility measures the probability of bridge components exceeding certain damage states at varying intensity measure (IM) levels of GMs. The *Cloud* analysis[55] considers the EDP-IM pairs to have a linear relationship in the logarithmic space, as shown in Eqn. (11). The associated dispersion corresponding to the linear regression is computed through Eqn. (12). The seismic fragility model is then estimated as a lognormal CDF by comparing seismic demands and capacities of bridge components, as shown in Eqn. (13).

$$log(\mu_{EDP}) = a + b \cdot log(IM) \quad (11)$$

$$\beta_{EDP} = \sqrt{\frac{1}{M-2}\sum_{1}^{M}[log(edp_i) - log(\mu_{EDP})]^2} \quad (12)$$

$$P[EDP \geq C|IM] = \Phi\left[\frac{log(\mu_{EDP}) - log(C)}{\sqrt{\beta_{EDP}^2 + \beta_C^2}}\right] \quad (13)$$

where *EDP* is the engineering demand parameter of interest, which is adopted as the drift ductility and park-Ang index[48] for the column, and peak displacement for the bearing; *C* is the corresponding seismic capacity model, as shown in Table S2; $\mu_{EDP}$ is the mean value of the seismic demand for the given *EDP*; *M* is the number of IM-EDP pairs available for analysis, and $edp_i$ indicates the $i^{th}$ seismic demand parameter obtained from the NRHAs. Column drift ductility equals its lateral displacement demand divided by the yielding displacement; the lateral displacement of the bearing can be directly extracted from the NRHAs. The Park-Ang damage index ($D_{PA}$) of the column is computed as:

$$D_{PA} = \frac{\Delta_{max}}{\Delta_u} + \psi_{PA}\frac{E_h}{F_y\Delta_u} \quad (14)$$

where $\Delta_m$ is column's maximum displacement demand; $\Delta_u$ is the displacement capacity corresponding to the same drift ductility at each damage state (Table S2); $\psi_{PA}$ controls the balance between the influence of peak deformation and cyclic damage accumulation, with a value of 0.05 used for the reinforced concrete structure[56]; $F_y$ denotes the yielding force, and $E_h$ is the total dissipated energy through hysteretic behavior.

## Normalized seismic loss ratio

The performance-based earthquake engineering framework[22] is utilized to compute the normalized annual seismic loss ratios of bridge components, as shown in Eq. (15).

$$SLR = \int_{IM}\sum_i d_i \times p_{ds,i}(IM) \times \left|\frac{d\lambda}{d(IM)}\right|d(IM) \quad (15)$$

where *SLR* denotes the normalized seismic loss ratio; $d_i$ is the damage ratio that represents the percentage in repair losses corresponding to damage state *i*, as listed in Table S2; $\lambda(IM)$ is characterized by the hazard curve, as shown in Fig. S9 for the considered bridge site; $p_{ds,i}(IM)$ is the in-state damage probability that can be computed by subtracting the consecutive damage state fragility values in Eqn. (13) at each IM level. More information regarding the *SLR* computation can be found in the Supplementary Material and elsewhere[57].

## References

1. Reduction, U. N. O. for D. R. *The Human Cost of Disasters: An Overview of the Last 20 Years (2000–2019)*. (2020).
2. Du, A., Wang, X., Xie, Y. & Dong, Y. Regional seismic risk and resilience assessment: Methodological development, applicability, and future research needs – An earthquake engineering perspective. *Reliab. Eng. Syst. Saf.* **233**, 109104 (2023).
3. Ning, C., Xie, Y., Burton, H. & Padgett, J. E. Enabling efficient regional seismic fragility assessment of multi-component bridge portfolios through Gaussian process regression and active learning. *Earthq. Eng. Struct. Dyn.* **53**, 2929–2949 (2024).
4. León, J. A., Ordaz, M., Haddad, E. & Araújo, I. F. Risk caused by the propagation of earthquake losses through the economy. *Nat. Commun.* **13**, 2908 (2022).
5. Silva-Lopez, R., Bhattacharjee, G., Poulos, A. & Baker, J. W. Commuter welfare-based probabilistic seismic risk assessment of regional road networks. *Reliab. Eng. Syst. Saf.* **227**, 108730 (2022).
6. Ceferino, L., Mitrani-Reiser, J., Kiremidjian, A., Deierlein, G. & Bambarén, C. Effective plans for hospital system response to earthquake emergencies. *Nat. Commun.* **11**, 4325 (2020).
7. Opabola, E. A. & Galasso, C. Informing disaster-risk management policies for education infrastructure using scenario-based recovery analyses. *Nat. Commun.* **15**, 325 (2024).
8. Koks, E. E. *et al.* A global multi-hazard risk analysis of road and railway infrastructure assets. *Nat. Commun.* **10**, 2677 (2019).
9. Cremen, G. & Galasso, C. Earthquake early warning: Recent advances and perspectives. *Earth-Science Rev.* **205**, 103184 (2020).
10. Allen, R. M., Gasparini, P., Kamigaichi, O. & Bose, M. The Status of Earthquake Early Warning around the World: An Introductory Overview. *Seismol. Res. Lett.* **80**, 682–693 (2009).
11. Strauss, J. A. & Allen, R. M. Benefits and costs of earthquake early warning. *Seismol. Res. Lett.* **87**, 765–772 (2016).
12. Given, D. *et al. Revised technical implementation plan for the ShakeAlert system—An earthquake early warning system for the West Coast of the United States*. (2018).
13. Hoshiba, M., Kamigaichi, O., Saito, M., Tsukada, S. & Hamada, N. Earthquake early warning starts nationwide in Japan. *EOS, Trans. Am. Geophys. union* **89**, 73–74 (2008).
14. Espinosa-Aranda, J. M. *et al.* Evolution of the Mexican seismic alert system (SASMEX). *Seismol. Res. Lett.* **80**, 694–706 (2009).
15. Zhang, H. *et al.* An Earthquake Early Warning System in Fujian, China. *Bull. Seismol. Soc. Am.* **106**, 755–765 (2016).
16. Field, E. H. *et al.* Uniform California earthquake rupture forecast, version 3 (UCERF3)—The time-independent model. *Bull. Seismol. Soc. Am.* **104**, 1122–1180 (2014).
17. Schiappapietra, E. & Douglas, J. Modelling the spatial correlation of earthquake ground motion: Insights from the literature, data from the 2016–2017 Central Italy earthquake sequence and ground-motion simulations. *Earth-Science Rev.* **203**, 103139 (2020).
18. Graves, R. *et al.* CyberShake: A Physics-Based Seismic Hazard Model for Southern California. *Pure Appl. Geophys.* **168**, 367–381 (2011).
19. Atkinson, G. M. Earthquake time histories compatible with the 2005 National building code of Canada uniform hazard spectrum. *Can. J. Civ. Eng.* **36**, 991–1000 (2009).
20. Rezaeian, S., Stewart, J. P., Luco, N. & Goulet, C. A. Findings from a decade of ground motion simulation validation research and a path forward. *Earthq. Spectra* **40**, 346–378 (2024).
21. Academies, N. *et al. Disaster resilience: A national imperative*. (National Academies Press, 2012).
22. Moehle, J. & Deierlein, G. G. A framework methodology for performance-based earthquake engineering. in *13th world conference on earthquake engineering* vol. 679 (WCEE Vancouver, 2004).
23. Cremen, G. & Baker, J. W. A Methodology for Evaluating Component-Level Loss Predictions of the FEMA P-58 Seismic Performance Assessment Procedure. *Earthq. Spectra* **35**, 193–210 (2019).
24. Zhang, H., Yu, D., Li, G. & Dong, Z. A real-time seismic damage prediction framework based on machine learning for earthquake early warning. *Earthq. Eng. Struct. Dyn.* **53**, 593–621 (2024).
25. Lu, X. & Guan, H. *Earthquake disaster simulation of civil infrastructures: From tall buildings to urban areas*. (Springer, 2017).
26. Baker, J. W. Efficient analytical fragility function fitting using dynamic structural analysis. *Earthq. Spectra* **31**, 579–599 (2015).
27. Çelebi, M. *Seismic instrumentation of buildings (with emphasis on federal buildings)*. https://citeseerx.ist.psu.edu/document?repid=rep1&type=pdf&doi=91db0804d2efb70befbba08e44a2bd8628d811c6 (2002).
28. Pei, S. *et al.* Shake-Table Testing of a Full-Scale 10-Story Resilient Mass Timber Building. *J. Struct. Eng.* **150**, 4024183 (2024).
29. Deierlein, G. G. & Zsarnóczay, A. State of the art in computational simulation for natural hazards engineering. *NHERI SimCenter* (2021).
30. Journeay, M., LeSueur, P., Chow, W., Wagner, C. & L. *Physical exposure to natural hazards in Canada: An overview of methods and findings*. https://ostrnrcan-dostrncan.canada.ca/handle/1845/134576 (2022) doi:10.4095/330012.
31. US Army Corps of Engineers Hydrologic Engineering Center. National Structure Inventory. https://www.hec.usace.army.mil/confluence/nsi (2023).
32. FHWA (Federal Highway Administration). National Bridge Inventory (NBI). https://www.fhwa.dot.gov/bridge/nbi.cfm (2024).
33. Mangalathu, S., Jeon, J.-S., Padgett, J. E. & DesRoches, R. ANCOVA-based grouping of bridge classes for seismic fragility assessment. *Eng. Struct.* **123**, 379–394 (2016).
34. Goda, K. & Atkinson, G. M. Seismic performance of wood-frame houses in south-western British Columbia. *Earthq. Eng. Struct.*

35. *Dyn.* **40**, 903–924 (2011).
36. Abarca, A., Monteiro, R. & O'Reilly, G. J. Seismic risk prioritisation schemes for reinforced concrete bridge portfolios. *Struct. Infrastruct. Eng.* **0**, 1–21 (2023).
37. Abarca, A., Monteiro, R., O'Reilly, G., Zuccolo, E. & Borzi, B. Evaluation of intensity measure performance in regional seismic risk assessment of reinforced concrete bridge inventories. *Struct. Infrastruct. Eng.* **19**, 760–778 (2023).
38. Ning, C. & Xie, Y. Convolutional variational autoencoder for ground motion classification and generation toward efficient seismic fragility assessment. *Comput. Civ. Infrastruct. Eng.* (2023) doi:10.1111/mice.13061.
39. Zhang, R. *et al.* Deep long short-term memory networks for nonlinear structural seismic response prediction. *Comput. Struct.* **220**, 55–68 (2019).
40. Li, Z. *et al.* Time history seismic response prediction of multiple homogeneous building structures using only one deep learning-based Structure Temporal Fusion Network. *Earthq. Eng. Struct. Dyn.* **53**, 4076–4098 (2024).
41. Kuo, P.-C. *et al.* GNN-LSTM-based fusion model for structural dynamic responses prediction. *Eng. Struct.* **306**, 117733 (2024).
42. Weiss, K., Khoshgoftaar, T. M. & Wang, D. A survey of transfer learning. *J. Big Data* **3**, 9 (2016).
43. Rethage, D., Pons, J. & Serra, X. A wavenet for speech denoising. in *2018 IEEE International Conference on Acoustics, Speech and Signal Processing (ICASSP)* 5069–5073 (IEEE, 2018).
44. Ning, C., Xie, Y. & Sun, L. LSTM, WaveNet, and 2D CNN for nonlinear time history prediction of seismic responses. *Eng. Struct.* **286**, 116083 (2023).
45. Lundberg, S. A unified approach to interpreting model predictions. *arXiv Prepr. arXiv1705.07874* (2017).
46. Mangalathu, S. & Jeon, J.-S. Regional Seismic Risk Assessment of Infrastructure Systems through Machine Learning: Active Learning Approach. *J. Struct. Eng.* **146**, 04020269 (2020).
47. Ancheta, T. D. *et al.* NGA-West2 database. *Earthq. Spectra* **30**, 989–1005 (2014).
48. Jalayer, F. Direct Probabilistic Seismic Anaysis: Implementing Non-Linear Dynamic Assessments. *Stanford University* (2003).
49. Shao, Y. & Xie, Y. Seismic risk assessment of highway bridges in western Canada under crustal, subcrustal, and subduction earthquakes. *Struct. Saf.* **108**, 102441 (2024).
50. U.S. Geological Survey. Earthquake lists, maps, and statisticsat. https://earthquake.usgs.gov/hazards/interactive (2020).
51. Naeim, F. & Kelly, J. M. *Design of seismic isolated structures: from theory to practice*. (Wiley& Sons, 1999).
52. Silva, P. F., Megally, S. & Seible, F. Seismic Performance of Sacrificial Exterior Shear Keys in Bridge Abutments. *Earthq. Spectra* **25**, 643–664 (2009).
53. Xie, Y. *et al.* Probabilistic models of abutment backfills for regional seismic assessment of highway bridges in California. *Eng. Struct.* **180**, 452–467 (2019).
54. Muthukumar, S. & DesRoches, R. A Hertz contact model with non-linear damping for pounding simulation. *Earthq. Eng. Struct. Dyn.* **35**, (2006).
55. Kingma, D. P. & Ba, J. Adam: A Method for Stochastic Optimization. *arXiv Prepr. arXiv1412.6980* (2014).
56. Jalayer, F., Ebrahimian, H., Miano, A., Manfredi, G. & Sezen, H. Analytical fragility assessment using unscaled ground motion records. *Earthq. Eng. Struct. Dyn.* **46**, 2639–2663 (2017).
57. Ghosh, J., Padgett, J. E. & Sánchez-Silva, M. Seismic Damage Accumulation in Highway Bridges in Earthquake-Prone Regions. *Earthq. Spectra* **31**, 115–135 (2015).
58. Ning, C. & Xie, Y. Risk-Based Optimal Design of Seismic Protective Devices for a Multicomponent Bridge System Using Parameterized Annual Repair Cost Ratio. *J. Struct. Eng.* **148**, (2022).

## Data availability

The data that support the findings of this study are available on GitHub (https://github.com/Ning-Civil/SPR-Net). All other data are available from the authors.

## Code availability

The codes are available on GitHub (https://github.com/Ning-Civil/SPR-Net).

## Acknowledgements

This research has been supported by the NSERC Discovery Grant of Canada under Funding No. RGPIN-2020-04156. Any opinions, findings, conclusions, or recommendations expressed in this paper are those of the authors and do not necessarily reflect the official views or policies of the funding agencies.

# Surrogate Structure-Specific Probabilistic Dynamic Responses of Bridge Portfolios using Deep Learning with Partial Information

Chunxiao Ning[1] | Yazhou Xie[1]

[1]Department of Civil Engineering, McGill University, Montreal, QC, Canada. Email: tim.xie@mcgill.ca

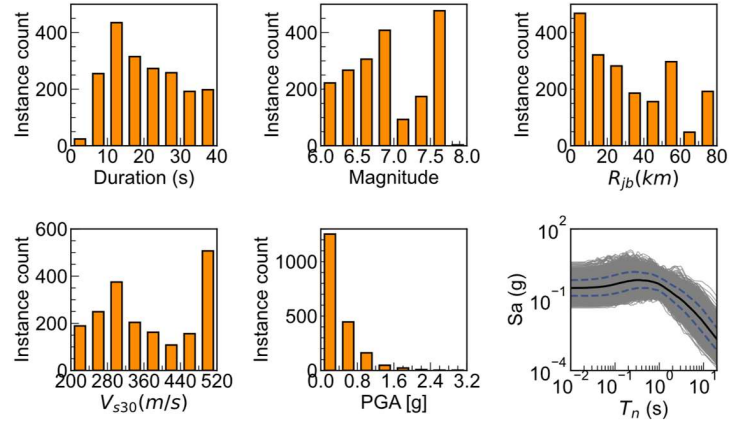

**Figure S1. Physical features of the selected ground motion dataset**

Fig. S1 shows statistical distributions of GMs' physical features, including earthquake magnitude, duration, peak ground acceleration (PGA), Joyner-Boore distance ($R_{jb}$), average shear wave velocity to a depth of 30 meters ($V_{S30}$) of recorded GM sites, and the acceleration response spectra ($S_a$) of all motions. Since bridges are considered to be geographically spread across the state of California, no site-specific target response spectrum is utilized for selecting the GMs. Instead, a general suite of 650 bi-directional GM records is selected from the NGA-West2 strong motion database[1]. The GMs are selected to have earthquake magnitudes $M_w$ larger than 6.0, the Joyner-Boore distance ($R_{jb}$) less than 80 km. These criteria eliminate small motion records that do not cause significant seismic damage. As a result, the selected GMs consist of 221 records from earthquake events in California and 429 motions recorded outside California. It is worth mentioning that all GMs are recorded from shallow crustal earthquakes in active tectonic regions, which somewhat resemble the seismicity in California. These GMs are further scaled twice with a random scaling factor ranging from 1 to 3 to generate more motions that are strong in intensity to cause significant nonlinearity and damage to bridge components. As such, a total suite of 1950 GM records is compiled.

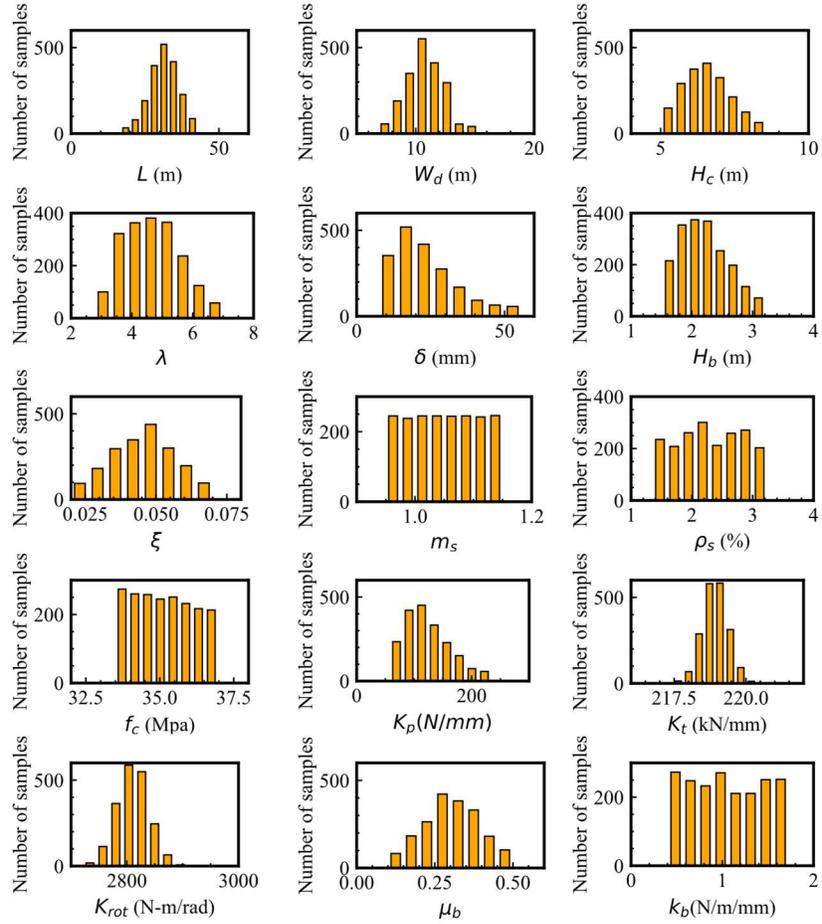

**Figure S2. Distributions of fifteen bridge parameters from bridge model sampling**

All bridges in the considered class are characterized by 15 independent modeling parameters, including span length $L(m)$, deck width $W_d(m)$, column height $H_c(m)$, column slenderness $\lambda$ defined as the height over diameter, abutment seat longitudinal gap length $\delta(mm)$, abutment backwall height $H_b(mm)$, damping ratio $\xi$, mass factor $m_s$, column longitudinal reinforcement ratio $\rho_s(\%)$, concrete strength $f_c(Mpa)$, abutment pile stiffness $K_p(N/mm)$, translational $K_t(kN/mm)$ and rotational stiffness $K_{rot}(Nm/rad)$ of column foundation, bearing friction coefficient $\mu_b$ and stiffness $k_b(N/m/mm)$. Fig. S2 shows the statitiscal distributions of these 15 modeling parameters from 1950 bridge model samples established for nonlinear dynamic response analyses.

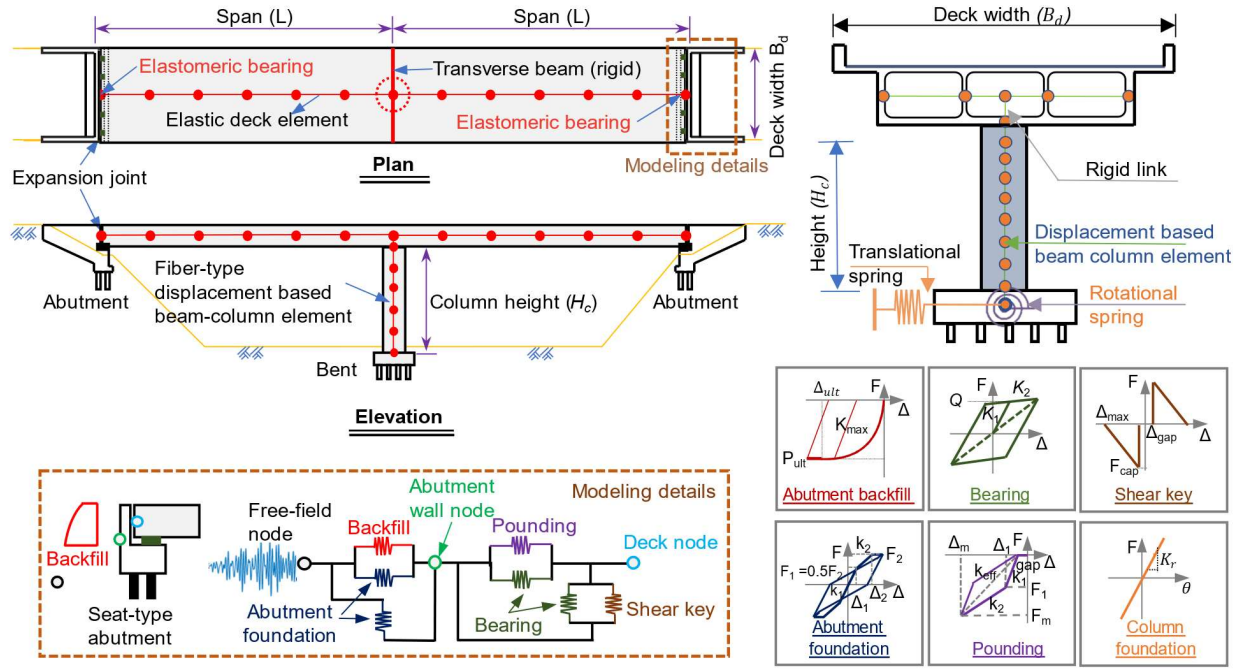

**Figure S3. High-fidelity numerical model of each bridge sample**

Fig. S3 shows the high-fidelity numerical model for the two-span, single-column, continuous concrete box-girder bridge designed before 1971 in California, United States. This bridge class represents a significant portion of the bridge inventory in the state of California[2,3]. The numerical models are developed in OpenSees to simulate the seismic responses of 1950 bridges in the same class. As shown in Fig. S3, the bridge deck is simulated using elastic beam elements with mass lumped along the centerline. The connections between the bridge deck and end abutments are preserved through rigid transverse beams. The columns are simulated using displacement-based beam-column elements with discretized fiber-defined sections. The *Concrete02* material is used to model the concrete, and material properties for the confined concrete are computed using Mander's model[4]. The *Steel02* material simulates the steel reinforcement in the column section[4]. The Zero-Length-Section element is added at the bottom of the column to model the strain penetration effect[5]. Elastic translational and rotational springs are used to model the column foundation, whereas a spring system is established to simulate the dynamic interplay among various abutment components, including the backfill, bearing, shear key, and pile foundation. A trilinear material is used to model the abutment pile foundation. The bilinear hardening material is used to simulate the nonlinear behavior of the bearing. The contact element developed by Muthukumar and DesRoches[6] is adopted to simulate the pounding effect between the deck and the abutment wall. The shear keys' force-displacement response is modeled using a trilinear curve based on the Caltrans-UCSD field experiments[7], and the seismic resistance of abutment backfills is simulated using a hyperbolic material[8].

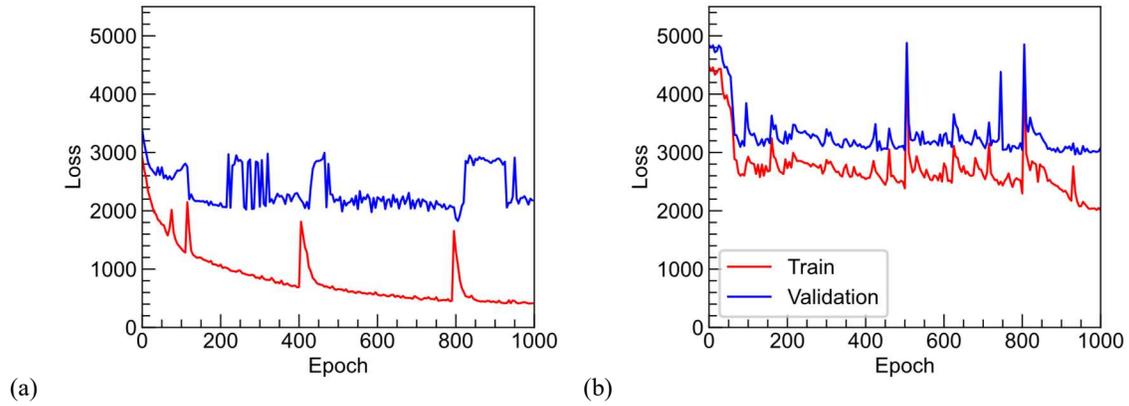

**Figure S4.** The training history of the (a) deterministic SPR-Net model and (b) LSTM model.

Fig. S4 shows the training history of the loss functions for the deterministic SPR-Net model and reference LSTM model. The deterministic SPR-Net takes GM sequences (Fig. S1) and values for the 15 bridge parameters (Fig. S2) as model inputs; the LSTM takes the GM sequences (Fig. S1) as the sole inputs. The loss function for both models is the mean squared error. The models used are those with the smallest validation losses, which are 1825.5 for the SPR-Net model at the $805^{th}$ epoch, and 2965.6 for the LSTM model at the $935^{th}$ epoch.

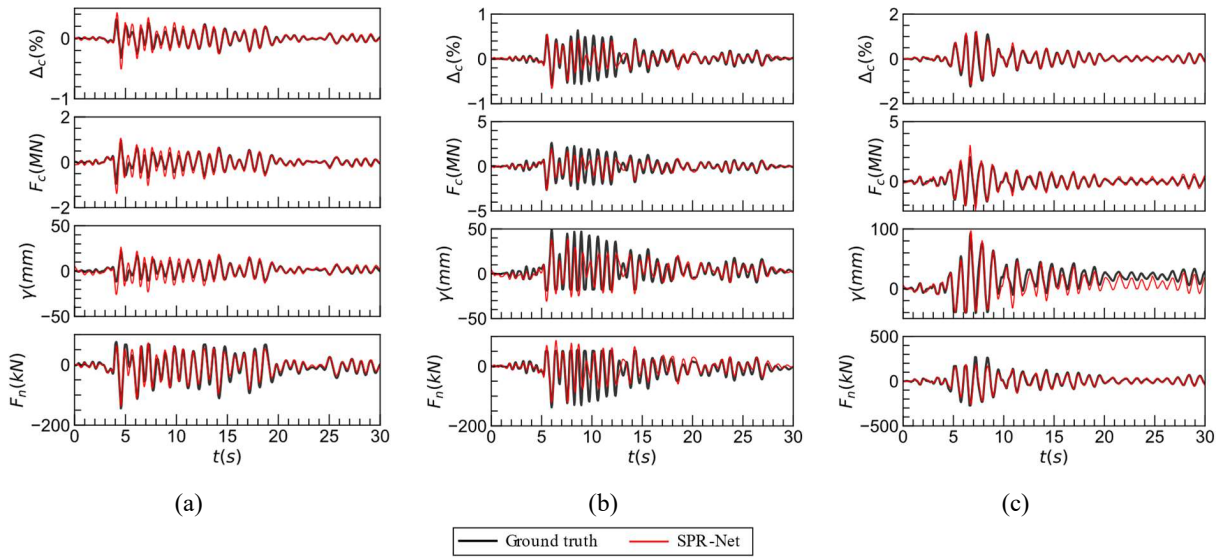

**Figure S5.** Seismic response predictions of the deterministic SPR-Net model for column peak drifts at (a) 16th, (b) 50th, and (c) 84th CDF percentiles

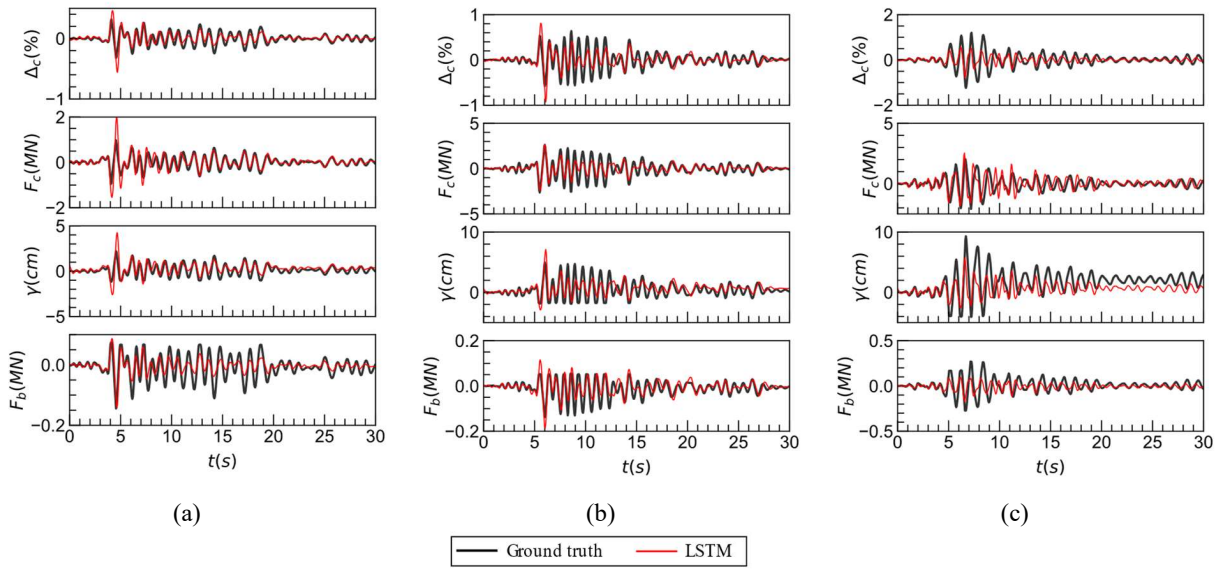

**Figure S6.** Seismic response predictions of the LSTM model for column peak drifts at (a) 16th, (b) 50th, and (c) 84th CDF percentiles

Figs. S5 and S6 compare the seismic response history predictions between the deterministic SPR-Net model and the LSTM model. Three cases from the testing dataset were selected for the comparison, corresponding to those at the 16th, 50th, and 84th CDF percentiles of the peak column drift ratios (i.e., Fig. 4(a)). Fig. S5 shows consistent time history predictions on the column and bearing responses from the deterministic SRP-Net model, while those from the LSTM model in Fig. S6 show substantial prediction discrepancies, particularly for the 50th and 84th percentile cases. Quantitative comparisons of these two models using a comprehensive set of evaluation metrics are provided in Fig. S7.

$$\mathcal{L}_{residual} = \frac{|y_{residual}^{pred} - y_{residual}^{true}|}{|y_{residual}^{true}|} \tag{S1}$$

$$\mathcal{L}_{peak} = \frac{\max|y^{true}| - \max|y^{pred}|}{\max|y^{true}|} \tag{S2}$$

$$\mathcal{A} = \frac{1}{l}\sum_{i}^{l} \frac{|y_i^{true} - y_i^{pred}|}{|y_i^{true}|} \tag{S3}$$

$$\mathcal{E} = \frac{\sum_i^l |y_i^{true}| - \sum_i^l |y_i^{pred}|}{\sum_i^l |y_i^{true}|} \tag{S4}$$

$$\mathcal{R}^2 = R^2[y^{pred}, y^{true}] \tag{S5}$$

$$\mathcal{N} = PDF\left\{\frac{y^{true} - y^{pred}}{max|y^{true}|}\right\} \tag{S6}$$

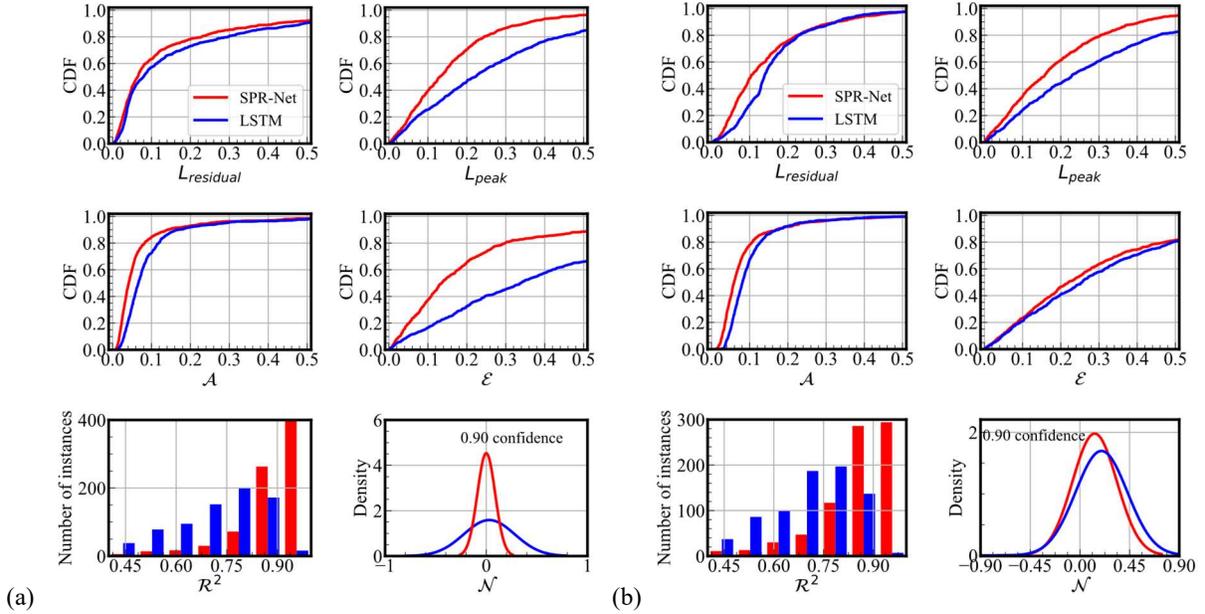

**Figure S7. Comparisons of the deterministic SPR-Net model and the LSTM model for (a) column drift and (b) the bearing displacement**

A comprehensive set of evaluation metrics is used to assess and compare the performance of the deterministic SPR-Net model. These metrics include the (1) relative errors in residual ($\mathcal{L}_{residual}$ in Eqn. S1) and peak responses ($\mathcal{L}_{peak}$ in Eqn. S2); (2) amplitude loss ($\mathcal{A}$ in Eqn. S3) that calculates the mean value of the amplitude differences over the entire $l$ time steps; (3) energy loss ($\mathcal{E}$ in Eqn. S4) that calculates the relative error in the time integration of absolute responses; (4) coefficient of determination ($\mathcal{R}^2$ in Eqn. S5) between predicted and true responses; and (5) probability density function (PDF) of the normalized error ($\mathcal{N}$ in Eqn. S6) at every time instant. In these equations, $y^{true}$, $y_i^{true}$, and $y_{residual}^{true}$ represent true responses over the entire time history, at time step $i$, and the residual value, respectively, and $y^{pred}$, $y_i^{pred}$, and $y_{residual}^{pred}$ represent those from predictions.

These six metrics evaluate the model performance both locally and globally. $\mathcal{L}_{residual}$, $\mathcal{L}_{peak}$ and $\mathcal{N}$ evaluate the prediction accuracy on specific time steps for local performance assessment, while $\mathcal{A}$, $\mathcal{E}$ and $\mathcal{R}^2$ capture the global quality of response predictions. Fig. S7 compares these six evaluation metrics between the deterministic SPR-Net model and LSTM model on the testing dataset for both column drift and bearing displacement responses. The CDF curves of $\mathcal{L}_{residual}$, $\mathcal{L}_{peak}$, $\mathcal{A}$, and $\mathcal{E}$ show that the SPR-Net model consistently outperforms the LSTM model in both local and global seismic response predictions. Additionally, $\mathcal{R}^2$ from the SPR-Net model are above 0.9 for most cases, much higher than the $\mathcal{R}^2$ values from the LSTM model. Fig. S7 shows that the developed SPR-Net model is more accurate and reliable than the reference LSTM model.

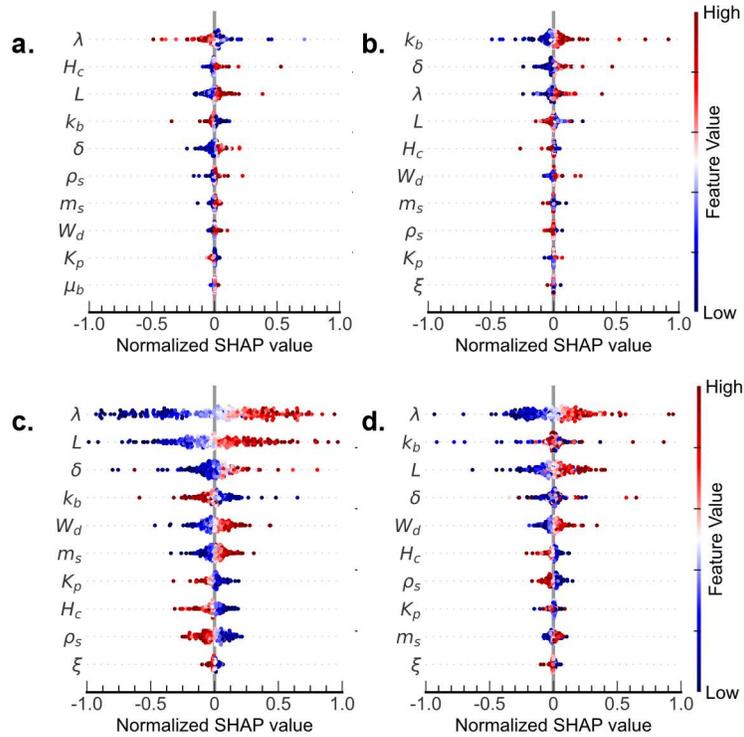

**Figure S8.** Shapley values of bridge parameters for (a) column energy dissipation, (b) bearing energy dissipation, (c) enclosed area of column deformation, and (d) enclosed area under bearing deformation

Fig. S8 shows the ranking in Shapley values of bridge parameters against column and bearing responses in terms of the dissipated energy (i.e., force-displacement products) and the enclosed deformation area along the entire time history. In addition to span length $L$, column slenderness $\lambda$, and abutment seat gap length $\delta$, other structural features, including bearing stiffness $k_b$, column height $H_c$, and deck width $W_d$, also show high rankings in Shapley values, demonstrating their substantial influences on the seismic responses of column and bearing. However, these other bridge attributes are not selected as the critical structural parameters for developing the probabilistic SPR-Net model. Bridge attributes such as $H_c$ have a high ranking against column dissipated energy; however, their Shapley value features a low dispersion, indicating that the impact of these parameters is limited to a small number of instances. Other parameters, such as bearing stiffness $k_b$, show considerable influences but are generally difficult to extract their values. Obtaining the values of these parameters requires either on-site testing or a thorough review of design documents. As a result, only features that are both influential and have their values easily assessable are retained, whereas all remaining parameters are treated as unknowns with a source of uncertainty through probabilistic seismic predictions. The results show that omitting these other features does not significantly degrade the performance of the probabilistic SPR-Net model.

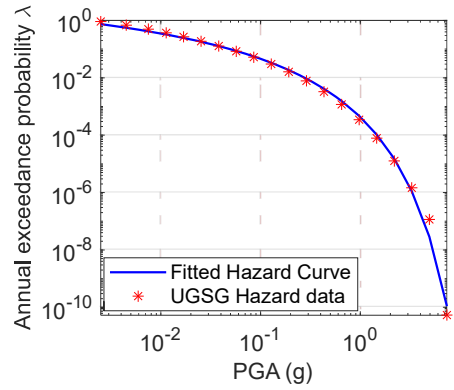

**Figure S9. Seismic hazard curve for the considered bridge site at Los Angeles**

The bridge for computing the normalized seismic loss ratio is considered to be physically located in the city of Los Angeles (33.98 N, −117.34 W), in California, United States. The associated hazard data for peak ground acceleration (PGA) can be extracted using the USGS unified hazard tool[9]. Eqn. S7 is further utilized to regress the hazard data against a continuous hazard curve.

$$\lambda(PGA) = \alpha \cdot \exp\left[\beta\left(\ln\left(\frac{PGA}{\gamma}\right)\right)^{-1}\right] \quad (S7)$$

The equation with $\alpha = 63.728, \beta = 42.475$, and $\gamma = 35.404$ fits the hazard data well, as shown in Fig. S9.

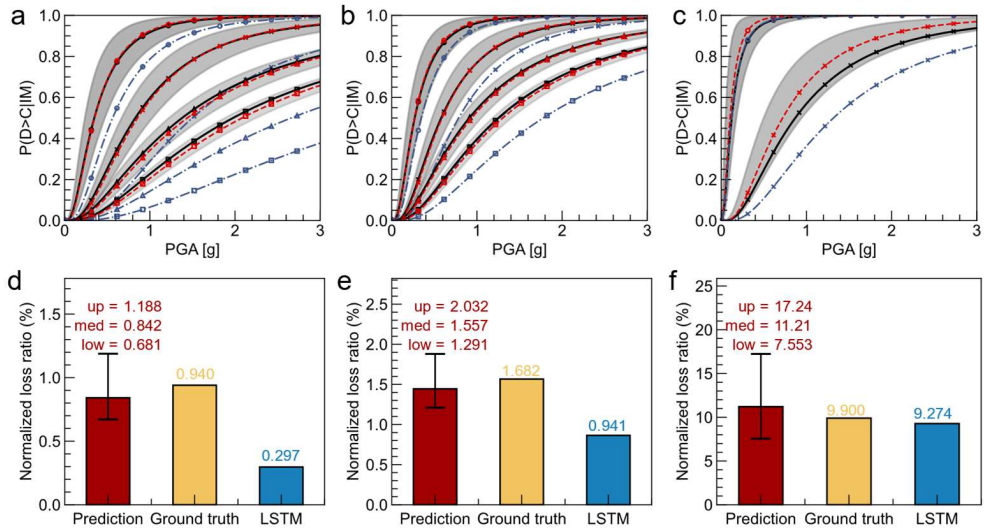

**Figure S10.** Comparisons of seismic fragility curves and seismic loss estimations for the medium bridge. (a) column fragility using drift ductility (b) column fragility using the Park-Ang index (c) bearing fragility (d) column loss ratio using drift ductility (e) column loss ratio using Park-Ang index and (f) bearing loss ratio.

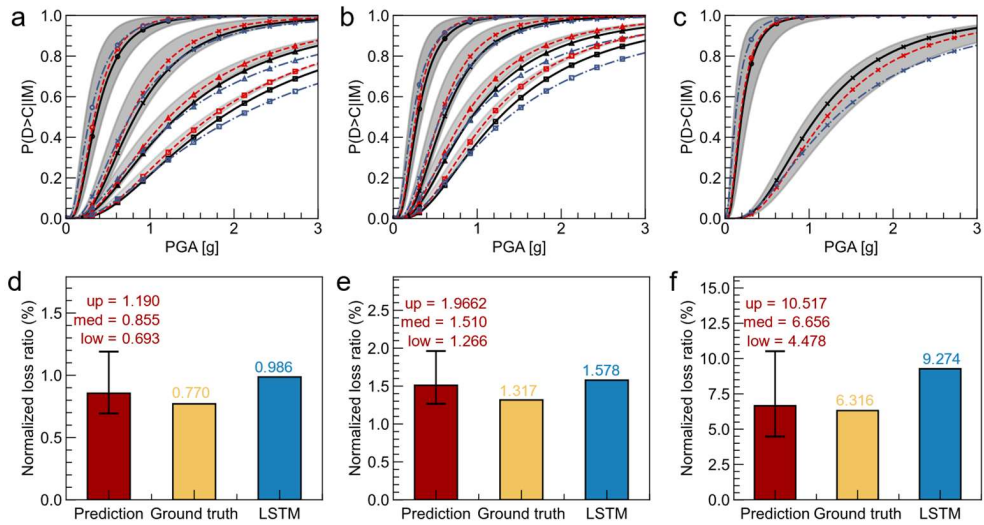

**Figure S11.** Comparisons of seismic fragility curves and seismic loss estimations for the stiff bridge. (a) column fragility using drift ductility (b) column fragility using the Park-Ang index (c) bearing fragility (d) column loss ratio using drift ductility (e) column loss ratio using Park-Ang index and (f) bearing loss ratio.

Fig. S10 compares the fragility and seismic loss results for the medium bridge that has a span length $L$=27.73 m, column slenderness ratio $\lambda$=5.60 and abutment seat gap length $\delta$=50.01 mm, resulting in a bridge natural period of 0.75 seconds. Fig. S10(a)-(c) show a matching median fragility and consistent bounds from the probabilistic predictions of the SPR-Net model. The LSTM model, however, significantly underestimates the column and bearing fragility for the medium bridge at higher damage states. Fig. S10(d)-(f) further compare the seismic repair cost ratios of the column and bearing for the medium bridge. No matter which EDP is considered for the column, Fig. S10(d) and S10(e) demonstrate that the SPR-Net predicts consistent normalized column loss ratios against the ground truth value. In contrast, the LSTM model offers a single-point prediction that is substantially lower in estimating column repair costs (i.e., 0.30% versus 0.94% for drift-ductility-based estimates).

Fig. S11 compares the same metrics for the stiff bridge that has a span length $L$=29.30 m, column slenderness ratio $\lambda$=3,69 and abutment seat gap length $\delta$=22.15 mm, resulting in a bridge natural period of 0.64 seconds. Fig. S11(a)-(c) show matching median fragility and consistent bounds from the probabilistic SPR-Net model. However, the LSTM model overestimates the fragilities at lower damage states and underestimates the ground truth against higher damage states. Results for the normalized seismic loss ratios also demonstrate reliable predictions from the SPR-Net model. Nevertheless, the LSTM model predicts much higher loss ratios for both the column (i.e., 0.99% versus 0.77% for drift-ductility-based estimates) and bearing (i.e., 9.27% versus 6.32%).

**Table S1. Statistics of bridge parameters considered for the bridge class**

| Parameters | Type | Mean | Standard deviation |
|---|---|---|---|
| Superstructure | | | |
|   Span length, $L_m(m)$ | LN | 31.78 | 8.74 |
|   Deck width, $W_D(m)$ | LN | 9.78 | 1.98 |
| Column | | | |
|   Height, $H_c(m)$ | LN | 6.63 | 0.87 |
|   Concrete strength, $f_c(MPa)$ | N | 29.03 | 3.59 |
|   Slenderness ratio, $\lambda$ | LN | 4.30 | 0.91 |
|   Rebar strength, $f_y(MPa)$ | LN | 465.0 | 37.30 |
|   Reinforcement ratio, $\rho_s(\%)$ | U | 2.25 | 0.52 |
| Foundation | | | |
|   Stiffness, $K_{ft}(kN/mm)$ | N | 245.78 | 105.08 |
|   Rotational stiffness, $K_{fr}(GN \cdot m/rad)$ | N | 6.8 | 1.1 |
| Abutment | | | |
|   Backwall height, $H_b(m)$ | LN | 2.19 | 0.44 |
|   Pile capacity, $K_p(kN/mm)$ | LN | 0.125 | 0.54 |
| Elastomeric bearing pad | | | |
|   Stiffness per deck width $k_b(kN/m^2)$ | LN | 908 | 327 |
|   Coefficient of friction for bearing pad $\mu_b$ | N | 0.30 | 0.10 |
| Expansion joint | | | |
|   Longitudinal gap (pounding), $\delta(mm)$ | LN | 23.5 | 12.5 |
|   Transverse gap (shear key), $\Delta(mm)$ | LN | 12.8 | 2.58 |
|   Acceleration for shear key capacity, $a_{sk}(g)$ | LN | 1 | 0.2 |
| Other parameters | | | |
|   Mass factor, $m_f$ | U | 1.05 | 0.06 |
|   Damping ratio, $\xi$ | N | 0.045 | 0.0125 |
|   Earthquake direction, $ED$ | B | - | - |

Note: N = normal; LN = lognormal; U = uniform; and B = Bernoulli distribution

Table S1 summarizes statistical distributions of 18 bridge parameters based on an extensive plan review of the bridge class in California to capture different sources of uncertainties[3]. 15 parameters are included as inputs to train the deterministic SPR-Net model, whereas steel strength $f_y$, abutment transverse gap length $\Delta$, and acceleration for shear key capacity $a_{sk}$, are excluded from modeling training. This is because (1) the steel strength $f_y$ of each bridge is found linear proportional to the corresponding concrete strength $f_c$ and (2) the abutment transverse gap $\Delta$ and the acceleration for shear key capacity $a_{sk}$ are not influential to bridge responses, particularly those along the longitudinal direciton.

**Table S2. Seismic capacity limit state model and damage ratios for bridge column and bearing**

| Component | EDP | Capacity Limit State Model | | | | | | | | Damage Ratios | | | |
|---|---|---|---|---|---|---|---|---|---|---|---|---|---|
| | | Median | | | | Dispersion | | | | | | | |
| | | S[1] | M[1] | E[1] | C[1] | S | M | E | C | S | M | E | C |
| Column | Drift ductility | 1.0 | 2.0 | 3.0 | 4.0 | | | | | 0.03 | 0.08 | 0.25 | 1.00 |
| | Park-Ang index | 1.0[2] | 1.0[2] | 1.0[2] | 1.0[2] | 0.25 | 0.25 | 0.47 | 0.47 | | | | |
| Bearing | Displacement (mm) | 25.4 | 101.6 | - | - | | | | | 0.50 | 1.00 | - | - |

Note: [1]S, M, E, and C represent slight, medium, extensive, and collapse damage states, respectively; [2]Park-Ang damage index considers 1.0 of reaching each damage state, where $\Delta_u$ in Eqn. 14 equals the drift ductility at each damage state (i.e., 1.0 – 4.0 from slight to complete damage) multiplied by the column yielding displacement.

Three EDPs are selected to assess the seismic fragility and loss ratios of the bridge column and bearing. The corresponding seismic capacity model and damage ratios are shown in Table S2. The column is considered to have four damage states as defined by HAZUS[10–12], and the bearing has two damage states[11]. The seismic capacities of each bridge component are assumed to follow lognormal distributions with median values listed in the table, and dispersions of 0.25 for slight and moderate damage and 0.47 for extensive and collapse damage. Damage ratios for these two components are adopted from the recommendation[13].

**References**
1. Ancheta, T. D. *et al.* NGA-West2 database. *Earthq. Spectra* **30**, 989–1005 (2014).
2. Mangalathu, S. & Jeon, J.-S. Regional Seismic Risk Assessment of Infrastructure Systems through Machine Learning: Active Learning Approach. *J. Struct. Eng.* **146**, 04020269 (2020).
3. Mangalathu, S., Jeon, J.-S., Padgett, J. E. & DesRoches, R. ANCOVA-based grouping of bridge classes for seismic fragility assessment. *Eng. Struct.* **123**, 379–394 (2016).
4. Filippou, F. C., Popov, E. P. & Bertero, V. V. Effects of bond deterioration on hysteretic behavior of reinforced concrete joint. *Earthq. Eng. Res. Center, Univ. California, Berkeley* 212 (1983).
5. Moridani, K. K. & Zarfam, P. Nonlinear Analysis of Reinforced Concrete Joints with Bond-Slip Effect Consideration in OpenSees. **3**, 362–367 (2013).
6. Muthukumar, S. & DesRoches, R. A Hertz contact model with non-linear damping for pounding simulation. *Earthq. Eng. Struct. Dyn.* **35**, (2006).
7. Silva, P. F., Megally, S. & Seible, F. Seismic Performance of Sacrificial Exterior Shear Keys in Bridge Abutments. *Earthq. Spectra* **25**, 643–664 (2009).
8. Xie, Y. *et al.* Probabilistic models of abutment backfills for regional seismic assessment of highway bridges in California. *Eng. Struct.* **180**, 452–467 (2019).
9. U.S. Geological Survey. Earthquake lists, maps, and statisticsat. https://earthquake.usgs.gov/hazards/interactive (2020).
10. FEMA. *Hazus -MH 2.1 Advanced Engineering Building Module (AEBM)*. (Federal Emergency Management Agency, 2021).
11. Ning, C., Xie, Y., Burton, H. & Padgett, J. E. Enabling efficient regional seismic fragility assessment of multi-component bridge portfolios through Gaussian process regression and active learning. *Earthq. Eng. Struct. Dyn.* **53**, 2929–2949 (2024).
12. Shao, Y. & Xie, Y. Seismic risk assessment of highway bridges in western Canada under crustal, subcrustal, and subduction earthquakes. *Struct. Saf.* **108**, 102441 (2024).
13. Ning, C. & Xie, Y. Risk-Based Optimal Design of Seismic Protective Devices for a Multicomponent Bridge System Using Parameterized Annual Repair Cost Ratio. *J. Struct. Eng.* **148**, (2022).